\documentclass[10pt,journal,compsoc]{IEEEtran}

\ifCLASSOPTIONcompsoc
  \usepackage[nocompress]{cite}
\else
  \usepackage{cite}
\fi

\usepackage{flushend}

\usepackage{amsmath}
\usepackage{array}
\usepackage{lipsum}
\usepackage{listings}

\usepackage{xspace}
\usepackage{blindtext}
\usepackage{hyperref}

\usepackage{multirow}

\newcommand{\EMPH}[1]{\emph{#1}}

\newcommand{\D}{$\Delta$\xspace}

\newcommand{\APPR}{\emph{DaMAT}\xspace}
\newcommand{\GomSpace}{GomSpace Luxembourg\xspace}

\newcommand{\LuxSpace}{LuxSpace\xspace}

\newcommand{\ESAIL}{ESAIL\xspace}

\newcommand{\LAUNCH}{on September 2020~\cite{ESAILlaunch}}

\newcommand{\SAIL}{\emph{\ESAIL}\xspace}

\newcommand{\PARAM}{\emph{LIBP}\xspace}

\newcommand{\CITSAIL}{~\cite{ESAIL}}

\newcommand{\UPDATED}[1]{\textcolor{black}{#1}}

\newcommand{\CHANGED}[1]{\textcolor{black}{#1}}
\newcommand{\NEWRES}[1]{#1}

\usepackage[backgroundcolor=white,bordercolor=blue,linecolor=blue]{todonotes}
\newcommand{\REVIEW}[2]{#2}

\newcommand{\TSEmin}[2]{#2}

\newcommand{\YAGO}{Yago Isasi Parache}

\newcommand{\EduardoSpace}{GomSpace Luxembourg\xspace}
\newcommand{\YagoSpace}{LuxSpace\xspace}

\newcommand{\ADCS}{\emph{\ESAIL-ADCS}\xspace}
\newcommand{\GPS}{\emph{\ESAIL-GPS}\xspace}
\newcommand{\PDHU}{\emph{\ESAIL-PDHU}\xspace}
\newcommand{\SVF}{\emph{\ESAIL-SVF}\xspace}

\def\BibTeX{{\rm B\kern-.05em{\sc i\kern-.025em b}\kern-.08em
    T\kern-.1667em\lower.7ex\hbox{E}\kern-.125emX}}

\newcommand\copyrighttext{%
  \footnotesize \textcopyright 2022 IEEE. Personal use of this material is permitted.
  Permission from IEEE must be obtained for all other uses, in any current or future
  media, including reprinting/republishing this material for advertising or promotional
  purposes, creating new collective works, for resale or redistribution to servers or
  lists, or reuse of any copyrighted component of this work in other works. 
  DOI: \href{TBA}{TBA}}
\newcommand\copyrightnotice{%
\begin{tikzpicture}[remember picture,overlay]
\node[anchor=south,yshift=10pt] at (current page.south) {\fbox{\parbox{\dimexpr\textwidth-\fboxsep-\fboxrule\relax}{\copyrighttext}}};
\end{tikzpicture}%
}
    
\begin{document}

\title{Data-driven Mutation Analysis for Cyber-Physical Systems}

\author{Enrico~Viganò, Oscar~Cornejo, Fabrizio~Pastore,~\IEEEmembership{Member,~IEEE,} and~Lionel~C.~Briand,~\IEEEmembership{Fellow,~IEEE}%

\IEEEcompsocitemizethanks{\IEEEcompsocthanksitem E. Viganò, O. Cornejo, F.
Pastore, and L. Briand are affiliated with SnT Centre for Security, Reliability and Trust, University of Luxembourg, Luxembourg.\protect\\
E-mail:\{enrico.vigano,oscar.cornejo,fabrizio.pastore,lionel.briand\}@uni.lu
\IEEEcompsocthanksitem L. Briand also holds a faculty appointment with school of EECS, University of Ottawa, Canada.}%
\thanks{Manuscript received April 01, 2022; revised September 09, 2022.}}

\IEEEtitleabstractindextext{
\begin{abstract}
Cyber-physical systems (CPSs) typically consist of a wide set of integrated, heterogeneous components; 
consequently, most of their critical failures relate to the  interoperability of such components. 
Unfortunately, most CPS test automation techniques are preliminary and industry still heavily relies on manual testing.
With potentially incomplete, manually-generated test suites, it is of paramount importance to assess their quality. 
Though mutation analysis has demonstrated to be an effective means to assess test suite quality in some specific contexts, we lack approaches for CPSs. Indeed, existing approaches do not target interoperability problems and cannot be executed in the presence of black-box or simulated components, a typical situation with CPSs.
In this paper, we introduce \emph{data-driven mutation analysis}, an approach that consists in assessing test suite quality by verifying if it detects interoperability faults simulated by mutating the data exchanged by software components. To this end, we describe a data-driven mutation analysis technique (\APPR) that automatically alters the data exchanged through data buffers. Our technique is driven by fault models in tabular form where engineers specify how to mutate data items by selecting and configuring a set of mutation operators. 
We have evaluated \APPR with CPSs in the space domain; specifically, the test suites for the software systems of a microsatellite and nanosatellites launched on orbit last year.
Our results show that the approach effectively detects  test suite shortcomings, is not affected by equivalent and redundant mutants, and entails acceptable costs.
\end{abstract}

}

\maketitle

\IEEEdisplaynontitleabstractindextext
\IEEEpeerreviewmaketitle

\copyrightnotice

\IEEEraisesectionheading{\section{Introduction}\label{sec:introduction}}

\IEEEPARstart{C}{yber} Physical Systems (CPSs) are heterogeneous systems that integrate computation, networking, and physical processes that are deeply interlaced~\cite{Khaitan:2015}. 
\CHANGED{In CPSs, conformance with requirements is verified 
through test cases 
executed at different development stages, based on available development artifacts~\cite{Roehm:2019}. In this paper, we focus on the identification of faults in the executable software to be deployed on the CPS, and thus target  software-in-the-loop (SIL) and hardware-in-the-loop (HIL) testing.}
 
 When software systems are large and integrate a diverse set of components, it is difficult to ensure that the test suite can detect any latent severe fault. To ensure \UPDATED{test suite quality}, standards for safety-critical software provide methodological guidance, for example through structural coverage adequacy; however, those strategies do not directly measure the fault detection capability of a test suite. A more direct solution to evaluate \UPDATED{test suite quality} is \emph{mutation analysis}~\cite{jia2010analysis,papadakis2019mutation}. It consists of automatically generating faulty software versions and %
 \UPDATED{computing} the mutation score, that is, the percentage of faulty software versions leading to a test failure.
 Mutation analysis is a good candidate to assess \UPDATED{test suite quality} because there is a strong association between high mutation scores and high fault revealing power for test suites~\cite{papadakis2018mutation}. 

Most mutation analysis techniques rely on the generation of faulty software through mutation operators that modify the software implementation (either the source or the executable code). 
Unfortunately, such techniques suffer from two major limitations, which are critical in the CPS context: (1) they cannot identify problems related to the interoperability of integrated components (integration testing) and (2) they can be applied only to components that can be executed in the development environment.
In CPSs, major problems typically arise because of the lack of  \emph{interoperability of integrated components}~\cite{Givehchi:2017,Jirkovsk:2017}, mainly due to the wide variety and heterogeneity of the technologies and standards adopted.
Also, there is  limited work on integration testing automation~\cite{Abbaspour:2015}, thus forcing companies to largely rely on manual approaches, which are error prone and likely to lead to incomplete test suites\TSEmin{3.1}{\footnote{In this paper, we rely on input domain partitioning to determine if a test suite is complete~\cite{Ammann:Offutt:2008}.}}. 
It is thus of fundamental importance to ensure the effectiveness of test suites with respect to detecting interoperability issues, for example by making sure test cases fully exercise the exchange of all possible data items and report failures when erroneous data is being exchanged by software components. For example, the test suite for the control software of a satellite shall identify failures due to components working with different measurement systems~\cite{MarsClimateOrbiter}.
Unfortunately, well known, code-driven mutation operators (e.g., the sufficient set~\cite{delamaro2014designing,delamaro2014experimental}) simulate algorithmic faults by introducing small changes into code and are thus unlikely to simulate interoperability problems resulting in \UPDATED{exchanges of erroneous data}.

The second limitation of code-driven mutation analysis approaches concerns \emph{the incapability of injecting faults into black-box components} whose implementation is not tested within the development environment (e.g., because it is simulated or executed on the target hardware).
For example, in a satellite system, such components include the control software of the Attitude Determination and Control System (ADCS), the GPS, and the Payload Data Handling Unit (PDHU). During SIL testing, the results generated by such components (e.g., the GPS position) are produced by a simulator. As for HIL testing, these components are directly executed on the target hardware and cannot be mutated, either because they are off-the-shelf components 
or 
to avoid damages potentially introduced by the mutation.

An alternative to code-driven mutation analysis approaches are model-based ones, which mutate models of the software under test (SUT). 
Unfortunately, existing approaches do not include strategies to simulate interoperability problems; also, their primary objective is to support test generation not the evaluation of test suites~\cite{He2011,Aichernig2015,Devroey2016,BELLI201625}. 
Furthermore, model-based test generation may not be cost-effective if detailed models of the system under test are not available---which is often the case, especially in early development stages---and lack key information required for testing (e.g., which telecommands shall trigger a state transition in a satellite system). 

To address the above-mentioned limitations, we propose \emph{data-driven mutation analysis}, 
a new mutation analysis paradigm
that alters the data exchanged by software components in a CPS to evaluate the capability of a test suite to detect interoperability faults. 
To this end, we present a technique, \emph{data-driven mutation analysis with tables} (\APPR),
to automate data-driven mutation analysis by relying on
a fault model that captures, for a specific set of components, both the characteristics of the data to mutate (e.g., the size and structure of the messages generated by the ADCS) and the types of fault that may affect such data (e.g., a  value out of the nominal range). The latter is formalized as a set of parameterizable mutation operators. 
Based on discussions with practitioners, to simplify adoption, we rely on fault models in tabular form where each row specifies, for a given data item, what mutation operator (along with its corresponding parameter values) to apply to which elements of the data item.
At runtime, \APPR modifies the data exchanged by components according to the provided fault model (e.g., replaces a nominal voltage value with a value out of the nominal range).

\REVIEW{A.5}{\APPR identifies test suite shortcomings that  consist of message types, software states, and input partitions not being exercised. 
It also addresses the lack of adequate test oracles, i.e.,
oracles capable of detecting observable failures caused by data (exchanged by SUT components) not being equivalent to the data assumed by test cases. 
We have defined three  analysis metrics enabling engineers to distinguish between such shortcomings and thus guiding the improvement of test suites; they are \emph{fault model coverage}, \emph{mutation operation coverage}, and \emph{covered mutation score} (see Section~\ref{sec:mutationAnalysisResults}).}

We performed an empirical evaluation of \APPR to determine the \UPDATED{effectiveness, feasibility, and applicability of data-driven mutation analysis for evaluating test suites.}
Our benchmark consists of software for CPSs in the space domain provided by our industry partners, which are 
the European Space Agency~\cite{ESA}, 
\GomSpace{}, a world-renowned manufacturer and supplier of nanosatellites, and 
\LuxSpace{}, a European developer of infrastructure products (e.g., microsatellites) and solutions for space.
More specifically, the benchmark includes (1) the on-board embedded 
{software system} for \SAIL{}\CITSAIL{}, a maritime microsatellite recently launched into space, and 
(2) a configuration library used in constellations of nanosatellites~\cite{satSurvey}. 
\UPDATED{Our empirical results show that \APPR (1)  successfully identifies different types of shortcomings in test suites, (2) prevents the introduction of equivalent and redundant mutants, and (3) is practically applicable in the CPS context.}

\REVIEW{A.1}{To summarize, our contributions include:
\begin{itemize}
    \item Data-driven mutation analysis, a new mutation analysis paradigm to assess how effectively a test suite detects interoperability faults. This paradigm is supported by an approach and a toolset detailed below. 
    \item A set of mutation operators that simulate interoperability problems in CPSs. Our mutation operators have been designed and validated with a group of space software experts that includes the heads of the software and testing departments of our industry partners, which are GomSpace, LuxSpace, and ESA.
    \item A methodology to define tabular fault models that characterize faults possibly affecting the data exchanged by CPS components at runtime.
    \item A set of adequacy metrics that enable engineers to distinguish between the kinds of problems potentially affecting their test suites, to better guide test suite improvements: data types not being exercised, input partitions not being covered, inadequate oracles, and application states not reached.
    \item \APPR, a technique that implements data-driven mutation analysis by mutating the data exchanged through data buffers in the SUT. \APPR simulates interoperability problems by applying a chosen set of mutation operators configured in a tabular fault model. It reports mutation analysis results using the newly introduced adequacy metrics.
    \item An empirical evaluation conducted with an industrial benchmark including space software currently on orbit.
    \item A replicability package~\cite{REPLICABILITY} and the source code of our toolset~\cite{DAMAT:Web} with a tutorial~\cite{DAMAT:Tutorial:Web}.
\end{itemize}}

This paper proceeds as follows. Section~\ref{sec:background}
describes background and related work. 
Section~\ref{sec:approach} describes the general principles of data-driven mutation analysis and our data-driven mutation technique, \APPR.
Section~\ref{sec:empirical} reports on the design and results of our empirical evaluation.
Section~\ref{sec:conclusion} concludes the paper.

\section{Background and Related Work}
\label{sec:background}

Data-driven mutation analysis evaluates the effectiveness of a test suite in detecting \textbf{interoperability faults}. The CPS literature reports on four different interoperability types~\cite{Givehchi:2017}: technical (which concerns communication protocols and  infrastructure), syntactic (which concerns data format), semantic (which concerns the exchanged information, that is, errors in the processing of exchanged data), and cross-domain interoperability (which concerns interaction through business process languages such as BPEL~\cite{BPEL}).
Technical and syntactic interoperability are provided by off-the-shelf hardware and libraries
(not tested by CPS developers) 
 while cross-domain interoperability concerns systems integrated in online services (e.g., energy plants) but is out of scope for the type of CPSs we target in this work, which are safety-critical CPSs like flight systems, robots, and automotive systems. In this paper, we thus focus on \emph{semantic interoperability} faults,
 that is, faults that affect CPS components integration and are triggered (i.e., lead to failures) in the presence of specific subsets of the data that might be exchanged by CPS components. We thus aim to ensure that a test suite fails when the data exchanged by CPS components is not the one specified by test cases (e.g., through simulator configurations).
Related work includes mutation analysis~\cite{jia2010analysis,papadakis2019mutation} and fault injection~\cite{natella2016assessing} techniques.

\textbf{Mutation analysis} concerns the automated generation of faulty software versions (i.e., mutants) through automated procedures called mutation operators~\cite{jia2010analysis,papadakis2019mutation}. The effectiveness of a test suite is measured by computing the mutation score, which is the percentage of mutants leading to failures when exercised by the test suite.

Mutation operators introduce syntactical changes into the code of the \UPDATED{SUT}. The  \emph{sufficient set of operators} is implemented by most mutation analysis toolsets~\cite{offutt1996experimental,rothermel1996experimental,andrews2005mutation,kintis2017detecting}. 
Unfortunately, these operators simulate faults concerning the implementation of algorithms (e.g., a wrong logical connector), which is usually tested in unit test suites that, by definition, do not exercise the communication among components, our target in this paper. 
Also, as stated in the Introduction, such operators cannot be used to generate faulty data with simulated or off-the-shelf components.
\emph{Higher-order} mutation analysis~\cite{harman2010manifesto}, which simply combines multiple operators, has the same limitations.

\REVIEW{B.16}{Recent work has introduced mutation operators for cyber-physical systems~\cite{zhu2018mutation}; they simulate low-level faults on hardware devices (e.g., by modifying the pin identifier of a general-purpose input/output integrated circuit) thus not addressing interoperability issues, our main focus in this paper. Further, they mutate the software implementation, thus presenting the same limitations as the approaches above.}

\UPDATED{Components integration is targeted by interface~\cite{delamaro2001interface}, integration~\cite{Grechanik:16}, contract-based~\cite{Jiang:ICSM:05}, and system-level mutation analysis~\cite{mateo2010mutation}. The former three assess the quality of integration test suites by introducing changes that concern function invocations 
(e.g., switch function arguments) and inter-procedural data-flow (e.g., alter assignments to variables returned to other components);
they can simulate integration faults in units integrated with API invocations but 
not interoperability problems concerning larger components communicating through channels (e.g., network).
System-level mutation relies on operators for GUI components, which are out of our scope, and  configuration files, by applying simple mutations, such as deleting a line of text, and are unlikely to lead to interoperability problems.}

\textbf{Fault injection techniques} simulate the effect of faults by altering, at runtime, the data processed by the \UPDATED{SUT}~\cite{natella2016assessing}. Faults are introduced according to a fault model that describes the type of fault to inject, the timing of the injection, and the part of the system targeted by the injection. Different from data-driven mutation analysis, fault injection techniques aim to stress the robustness of the software,  
not assess the quality of its test suites.

Faults affecting components' communication, 
CPU, or memory can be simulated by performing bit flips 
\cite{tsai1999stress,barton1990fault,han1995doctor,dawson1996testing}.
Communication faults are simulated also by duplicating or deleting packets, altering their sequence, or introducing incorrect identifiers, checksums, or counters~\cite{di2015evolutionary,di2015generating}.
Faults affecting signals can be simulated by shifting the signal or increasing the number of signal segments~\cite{Matinnejad19}.
The largest set of faults affecting data exchanged through files or byte streams is simulated by Peach~\cite{PeachFuzzer}, which includes also protocol-specific fault injection procedures such as replacing host names with randomly generated ones.
In general, although existing techniques may simulate a large set of faults they do not cover all the CPS interoperability faults (see Section~\ref{sec:faultModelStructure}).

Approaches performing fault injections other than bit flips require a model of the data to modify.
The modelling formalisms adopted for this purpose are grammars~\cite{ghosh1998testing,Godefroid:GrammarBasedFuzzying:2008,godefroid2012sage,bounimova2013billions}, UML class diagrams~\cite{di2015evolutionary,di2015generating}, or block models~\cite{pham2016model,PeachFuzzer}.
Grammars are used to model textual data (e.g., XML), which is seldom exchanged by CPS components because of parsing cost. 
Block models enable specifying the representation to be used for consecutive blocks of bytes, which makes them applicable to a large set of systems; however, existing block model formalisms rely on the XML format, which is expensive to process and thus not usable with real-time systems~\cite{pham2016model,PeachFuzzer}.
The UML class diagram is a formalism that
enables the specification of complex data structures and 
data dependencies 
\cite{di2015evolutionary,di2015generating}; however, it requires loading the data as UML class diagram instances, which is too expensive for real-time systems. 

\TSEmin{3.2}{Bai et al. introduced a set of fault injection operators that alter the data generated by the simulators used in CPS testing~\cite{Bai:07}. They aim to reduce test input selection costs and do not target mutation analysis; indeed, their operators enable engineers to test exceptional scenarios by reusing simulator configurations for nominal cases. Different from us, they support robustness testing, not test suite assessment; further, their approach does not rely on any data modeling solution and thus requires the re-implementation of each mutation operator for each SUT, an error-prone activity.} 

To summarize, the modification of the data exchanged by software components enables the simulation of communication and, therefore, semantic interoperability faults.
Test suites can thus be assessed by relying on fault injection techniques to mutate data. 
However, existing fault injection techniques do not target mutation analysis; consequently, we lack methods for the specification of fault models and metrics for the assessment of test suites. 
Also, a larger set of procedures for the modification of data is needed.
Finally, block models can effectively capture the structure of the data to modify but formalisms not relying on XML are needed. Our paper addresses such limitations.

\section{Approach}
\label{sec:approach}

\begin{figure}[tb]
	\centering
		\includegraphics[width=8.4cm]{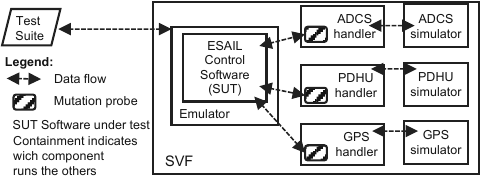}
		\caption{\CHANGED{Data mutation probes integrated into \ESAIL.}}
		\label{fig:appr:mutateProbesInserted}
	\end{figure}

\begin{figure}[tb]
	\centering
		\includegraphics[width=8cm]{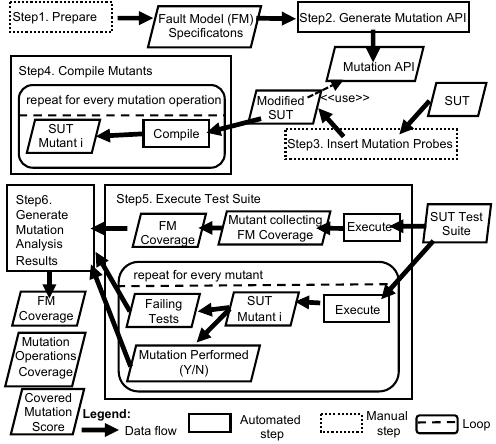}
		\caption{The \APPR process.}
		\label{fig:appr:approach}
	\end{figure}

\TSEmin{3.3}{In this section we provide an overview of \APPR (Section~\ref{sec:approach:overview}), followed by a description of its fault model (Section~\ref{sec:faultModelStructure}) and  each of its steps~(Sections~\ref{sec:methodology} to \ref{sec:mutationAnalysisResults}).}

\subsection{Overview}
\label{sec:approach:overview}

 Data-driven mutation analysis aims to evaluate the effectiveness of a test suite in detecting semantic interoperability \UPDATED{faults}. 
It is achieved by modifying (i.e., mutating) the data exchanged by CPS components. It generates \emph{mutated data} that is representative of what might be observed at runtime in the presence of a component that behaves differently than expected in the test case. 
\REVIEW{B.4}{The mutated data shall lead to different execution behavior than the original data to trigger test failures.
}
For these reasons, data mutation is driven by a fault model specified by the engineers based on domain knowledge.

Although different types of fault models might be envisioned,
in this paper we propose a technique (\emph{data-driven mutation analysis with tables}, \APPR),
which automates data-driven mutation analysis by relying on
a tabular \CHANGED{block model}, itself tailored to the \UPDATED{SUT} through predefined mutation operators.
To concretely perform data mutation at runtime, \APPR relies on a set of \emph{mutation probes} that shall be integrated by software engineers into the software layer that handles the communication between components. The runtime behaviour of mutation probes (i.e, what data shall be mutated and how) is driven by the fault model. Thus, \APPR can automatically generate the implementation of mutation probes from the provided fault model.
Depending on the CPS, probes might be inserted either into the \UPDATED{SUT}, into the simulator infrastructure, or both. 
Figure~\ref{fig:appr:mutateProbesInserted} shows the architecture of the \ESAIL satellite system (one of the subjects considered in our empirical evaluation) with mutation probes integrated into the SVF\footnote{Software Validation Facility~\cite{Isasi2019}; it usually includes one or more simulators, an emulator to run the code compiled for the target hardware, and test harnesses.} functions that handle communication with external components (PDHU, GPS, and ADCS in this case).

\APPR works in six steps, which are shown in Figure~\ref{fig:appr:approach}. 
In Step 1, based on the provided methodology and predefined mutation operators, the engineer prepares a fault model specification tailored to the SUT.
In Step 2, \APPR generates a mutation API with the functions that modify the data according to the provided fault model.
In Step 3, the engineer modifies the \UPDATED{SUT} by introducing mutation probes (i.e., invocations to the mutation API) into it.
In Step 4, \APPR generates and compiles mutants. 
Since the \APPR mutation operators may generate mutated data by applying multiple mutation procedures, \APPR may generate several mutants, one for each \UPDATED{mutation operation (i.e., a mutation procedure configured for a data item, according to our terminology, see Section~\ref{sec:mutantsGeneration})}.
In Step 5, \APPR executes the test suite with all the mutants including a mutant (i.e., the coverage mutant) which does not  modify the data but traces the coverage of the fault model (\emph{FM coverage} in Figure~\ref{fig:appr:approach}).
In Step 6, \APPR generates mutation analysis results.

\TSEmin{1.1}{The manual steps of \APPR (i.e., Step 1 and Step 3) cannot be automated because they require information that cannot be extracted automatically. The definition of the fault model in Step 1 requires a data model of the software, including valid and invalid input partitions, based on software requirement specification documents, a task that cannot be automated (see Section~\ref{sec:methodology}). Step 3 requires the identification of program statements handling the communication between components, which also cannot be automated (see Section~\ref{sec:generateAPI}).}

\subsection{Fault Model Structure}
\label{sec:faultModelStructure}

\newcommand{\MINIPM}{2.5cm}
\newcommand{\MINIPW}{11.5cm}
\begin{table*}
\caption{Data-driven mutation operators.}
\label{table:operators}
\scriptsize
\begin{tabular}{|p{17mm}|p{10mm}|p{2.5cm}|p{11.5cm}|}
\hline
\textbf{Fault Class}&\textbf{Types}&\textbf{Parameters}&\textbf{Description}\\
\hline
Value above threshold (VAT)&
I,L,F,D,H
&
\begin{minipage}{\MINIPM}
T: threshold\\
\D: delta, difference with respect to threshold\\
\end{minipage}
&
\begin{minipage}{\MINIPW}
Replaces the current value with a value above the threshold T for a delta (\D). It simulates a value that is out of the nominal case and shall trigger a response from the system that shall be verified by the test case (e.g., the system may continue working but an alarm shall be triggered). Not applied if the value is already above the threshold.

\EMPH{Mutation procedure:}{\tiny 
\[
v' =  
    \begin{cases}
      (T+\Delta)    & \mathit{if} v \le T\\
      v    & \mathit{otherwise}\\
    \end{cases}       
\]
}
\end{minipage}
\\

\hline
Value below threshold (VBT)&
I,L,F,D,H
&
\begin{minipage}{\MINIPM}
T: threshold\\
\D: delta, difference with respect to threshold\\
\end{minipage}
&
\begin{minipage}{\MINIPW}
Replaces the current value with a value below the threshold T for a delta (\D). It simulates a value that is out of the nominal case and shall trigger a response from the system that shall be verified by the test case (e.g., the system may continue working but an alarm shall be triggered). Not applied if the value is already below the threshold.

\EMPH{Mutation procedure:}{\tiny 
\[
v' =  
    \begin{cases}
      (T-\Delta)    & \mathit{if} v \ge T\\
      v    & \mathit{otherwise}\\
    \end{cases}       
\]}
\end{minipage}
\\

\hline
Value out of range (VOR)&
I,L,F,D,H
&
\begin{minipage}{\MINIPM}
MIN: minimum valid value\\
MAX: maximum valid value\\
\D: delta, difference with respect to minimum/maximum valid value
\end{minipage}
&
\begin{minipage}{\MINIPW}
Replaces the current value with a value out of the range $[MIN;MAX]$. It simulates a value that is out of the nominal range and shall trigger a response from the system that shall be verified by the test case (e.g., the system may continue working but an alarm shall be triggered). Not applied if the value is already out of range.
\CHANGED{This was inspired by the \emph{ARBC} operator~\cite{di2015generating}; however, \APPR enables engineers to explicitly specify the delta.}

\begin{minipage}{\MINIPW/2}

\EMPH{Mutation procedure 1:} {\tiny
\[
v' =  
    \begin{cases}
      (MIN-\Delta)    & \mathit{if} MIN \le v \le MAX\\
      v    & \mathit{otherwise}\\
    \end{cases}       
\]
}
\end{minipage}
\begin{minipage}{\MINIPW/2}
\EMPH{Mutation procedure 2:} {\tiny
\[
v' =  
    \begin{cases}
     (MAX+\Delta)    & \mathit{if} MIN \le v \le MAX\\
     v    & \mathit{otherwise}\\
    \end{cases}       
\]
}
\end{minipage}

\end{minipage}
\\

\hline
Bit flip (BF)&
B
&
\begin{minipage}{\MINIPM}
MIN: lower bit\\
MAX: higher bit\\
STATE: mutate only if the bit is in the given state (i.e., 0 or 1). \\
VALUE: number of bits to mutate\\
\end{minipage}
&
\begin{minipage}{\MINIPW}
A number of bits randomly chosen in the positions between MIN and MAX (included) are flipped.
If STATE is specified, the mutation is applied only if  the bit is in the specified state; the value $-1$ indicates that any state shall be considered for mutation. Parameter VALUE specifies the number of bits to mutate.
\CHANGED{This was inspired by the \emph{BitFlipperMutator} operator~\cite{PeachFuzzer}; however, \APPR introduces the STATE parameter, which is not supported by related work.}

\EMPH{Mutation procedure:} the operator flips VALUE randomly selected bits if they are in the specified state.

\end{minipage}
\\

\hline
Invalid numeric value (INV)&
I,L,F,D,H
&
\begin{minipage}{\MINIPM}
MIN: lower valid value\\
MAX: higher valid value\\
\end{minipage}
&
\begin{minipage}{\MINIPW}
Replace the current value with a mutated value that is legal (i.e., in the specified range) but different than current value. It simulates the exchange of data that is not consistent with the state of the system. 
\CHANGED{It matches the \emph{ARR} operator~\cite{di2015generating}.}

\EMPH{Mutation procedure:} replace the current value with a different value randomly sampled in the specified range.
\end{minipage}
\\

\hline
Illegal Value (IV)
&
I,L,F,D,H
&
\begin{minipage}{\MINIPM}
VALUE: illegal value that is observed\\
\end{minipage}
&
\begin{minipage}{\MINIPW}
Replace the current value with a value that is equal to the parameter \emph{VALUE}. 
\CHANGED{It matches the \emph{ValidValuesMutator} operator~\cite{PeachFuzzer}.}

\EMPH{Mutation procedure:} {\tiny
\[
v' =  
    \begin{cases}
     \mathit{VALUE}    & \mathit{if} v \ne \mathit{VALUE}\\
     v    & \mathit{otherwise}\\
    \end{cases}       
\]
}
\end{minipage}
\\

\hline
Anomalous Signal Amplitude (ASA)
&
I,L,F,D,H
&
\begin{minipage}{\MINIPM}
T: change point\\
\D: delta, value to add/remove\\
VALUE: value to multiply\\
\end{minipage}
&
\begin{minipage}{\MINIPW}
The mutated value is derived by amplifying the observed value by a factor \emph{VALUE} and by adding/removing a constant value \D from it. It is used to either amplify or reduce a signal in a constant manner to simulate unusual signals. The parameter \emph{T} indicates the observed value below which instead of adding  we subtract.

\EMPH{Mutation procedure:}{\tiny 
\[
v' =  
    \begin{cases}
     T+(  (T-v)*\mathit{VALUE}  ) + \Delta    & \mathit{if}\ v \ge T\\
     T - (  (v-T)*\mathit{VALUE}  ) - \Delta   & \mathit{if}\ v < T
    \end{cases}       
\]
}
\end{minipage}
\\

\hline
Signal Shift (SS)
&
I,L,F,D,H
&
\begin{minipage}{\MINIPM}
\D: delta, value by which the signal should be shifted\\
\end{minipage}
&
\begin{minipage}{\MINIPW}
The mutated value is derived by adding a value \D to the observed value. It simulates an anomalous shift in the signal.
\CHANGED{This was inspired by work on signal mutation~\cite{Matinnejad19}; however, \APPR also enables engineers to rely on SS to increment (or decrement) counters and identifiers.}

\EMPH{Mutation procedure:} 
$v' = v + \Delta$
\end{minipage}
\\

\hline
Hold Value (HV)
&
\begin{minipage}{\MINIPW}
I,L,F,D,H
\end{minipage}
&
\begin{minipage}{\MINIPM}
V: number of times to repeat the same value\\
\end{minipage}
&
\begin{minipage}{\MINIPW}
This operator keeps repeating an observed value for $V$ times. It emulates a constant signal replacing a signal supposed to vary.

\EMPH{Mutation procedure:} {\tiny 
\[
v' =  
    \begin{cases}
     \mathit{previous}\  v'   & \mathit{if}\ \mathit{counter} \le V\\
     v  & \mathit{otherwise}\
    \end{cases}       
\]
}
\end{minipage}
\\

\hline
Fix value above threshold (FVAT)&
I,L,F,D,H
&
\begin{minipage}{\MINIPM}
T: threshold\\
\D: delta, difference with respect to threshold\\
\end{minipage}
&
\begin{minipage}{\MINIPW}
It is the complement of VAT and implements the same mutation procedure as VBT but we named it differently because it has a different purpose. Indeed, it is used to verify that test cases exercising exceptional cases are verified correctly. In the presence of a value above the threshold, it replaces the current value with a value below the threshold T for a delta \D.

\EMPH{Mutation procedure:} {\tiny
\[
v' =  
    \begin{cases}
     (T-\Delta)    & \mathit{if} v > T\\
     v    & \mathit{otherwise}\\
    \end{cases}       
\]
}
\end{minipage}
\\

\hline
Fix value below threshold (FVBT)&
I,L,F,D,H
&
\begin{minipage}{\MINIPM}
T: threshold\\
\D: delta, difference with respect to threshold\\
\end{minipage}
&
\begin{minipage}{\MINIPW}
It is the counterpart of FVAT for the operator VBT.

\EMPH{Mutation procedure:} {\tiny
\[
v' =  
    \begin{cases}
     (T+\Delta)    & \mathit{if} v < T\\
     v    & \mathit{otherwise}\\
    \end{cases}       
\]
}
\end{minipage}
\\

\hline
Fix value out of range (FVOR)&
I,L,F,D,H
&
\begin{minipage}{\MINIPM}
MIN: minimum valid value\\
MAX: maximum valid value\\
\end{minipage}
&
\begin{minipage}{\MINIPW}
It is the complement of VOR and implements the same mutation procedure as INV but we named it differently because it has a different purpose. Indeed, it is used to verify that test cases exercising exceptional cases are verified correctly.

\EMPH{Mutation procedure:} {\tiny
\[
v' =  
    \begin{cases}
     v    & \mathit{if} MIN \le v \le MAX\\\\
     \mathit{random(MIN,MAX)}    & \mathit{otherwise}\\
    \end{cases}       
\]
}
\end{minipage}
\\

\hline
\end{tabular}
\\
\textbf{Legend:} I: INT, L: LONG INT, F: FLOAT, D: DOUBLE, B: BIN, H: HEX
\end{table*}%

The \APPR fault model enables the specification of the format of the data exchanged between components along with the type of faults that may affect such data. 
In this paper, we refer to the data exchanged by two components as \emph{message}; also, each CPS component may generate or receive different \emph{message types}.
For a single CPS, more than one fault model can be specified. For example, in the case of \ESAIL{} we have defined one fault model for every message type that could be exchanged by the three components under test (i.e., ADCS, PDHU, and GPS). In total, for \ESAIL, we have 14 fault models, 10 for the communication concerning ADCS (we have 10 different message types), 3 for PDHU, and 1 for GPS.
\REVIEW{B.6}{Our methodology for defining the SUT fault models is described in Section~\ref{sec:methodology}.}

\begin{figure}
	\centering
		\includegraphics[width=8.4cm]{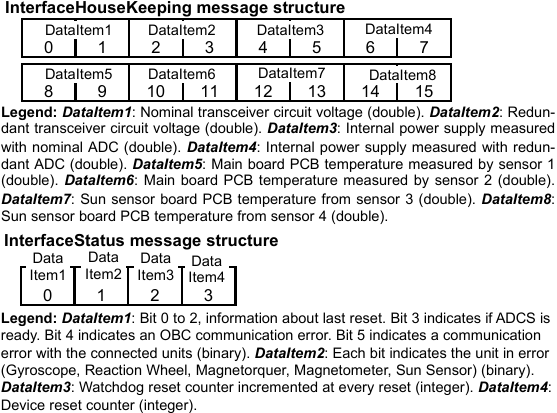}
		\caption{Structure of data buffers in \ESAIL.}
		\label{fig:appr:bufferStructure}
	\end{figure}

The \APPR fault model enables the modelling of data that is exchanged through a specific data structure: the data buffer. This was decided because it is a simple and widely adopted data structure for data exchanges between components in CPS. Also, more complex data structures (e.g., hierarchical ones like trees) are often flattened into data buffers in order to be exchanged by different components (e.g., through the network). When the CPS software is implemented in C or C++ (common CPS development languages) data buffers are implemented as arrays. Figure~\ref{fig:appr:bufferStructure} shows three block diagrams representing (part of) the buffer structure used to exchange messages of type InterfaceHouseKeeping and InterfaceStatus in \ESAIL.

A data buffer is characterized by a \emph{unit size} that specifies the dimension, in bytes, of the single cell of the underlying array and a \emph{buffer size}, which specifies the total number of units belonging to the buffer. Each data buffer can contain one or more \emph{data items}; the size of data items may vary as they may span over multiple units. Also, each data item is interpreted by the CPS software according to a specific \emph{representation} (e.g., integer, double, etc.). 
In \ESAIL, the unit size is one byte and the data items may span over one or two buffer units (see Figure~\ref{fig:appr:bufferStructure}). 

The \APPR fault model enables engineers to specify (1) the \emph{position} of each data item in the buffer, (2) their \emph{span}, and (3) their \emph{representation type}. Our current implementation supports six data representation types: int, long int, float, double, bin (i.e., data that should be treated in its binary form), hex (i.e., data that should be treated as hexadecimal).
Further, for each data item, \APPR enables engineers to specify one or more data faults using the mutation operator identifiers. %
For each operator, the engineer 
shall provide values for the required configuration parameters
\REVIEW{B.6}{(e.g., the nominal range for a numeric data item). 
}

Table~\ref{table:operators} provides the list of mutation operators included in \APPR along with their description \REVIEW{B.8}{and a description of their configuration parameters}. The \APPR mutation operators generate \emph{mutated data item instances} through one or more \emph{mutation procedures}, which are the functions that generate a mutated data item instance given a correct data item instance observed at runtime. For example, the \emph{VAT} operator includes only one mutation procedure (i.e., setting the current value above the threshold) while the \emph{VOR} operator includes two mutation procedures, which are
(1) replacing the current value with a value above the specified valid range and (2) replacing the current value with a value below the valid range.
\REVIEW{A.3}{The operators VOR, BF, INV, and SS have been inspired by related work~\cite{di2015generating,PeachFuzzer,Matinnejad19}; instead, the operators VAT, VBT, FVAT, FVBT, FVOR, IV, ASA,  and HV
are a contribution of this paper and were conceptualised from discussions with engineers having leading roles in ESA, GomSpace, and LuxSpace, our partners in this research\footnote{Our discussions involved the Head of Software for \ESAIL, the Mission Lead for GomSpace Luxembourg, the Head of the Flight Software Systems Section at ESA ESTEC, and several software engineers working with GomSpace, LuxSpace, and ESA.}.}

\TSEmin{1.2, 2.1}{Most of our operators generate mutated data values that are deterministic. Such design choice aims to (1) maximize the likelihood of reproducing test outcomes, which is necessary for debugging, and (2) maximize the likelihood of altering software behaviour and thus causing a test failure, which is necessary to avoid false alarms.
The operators generating mutated data values above or below thresholds (i.e., VAT, VBT, VOR, FVAT, FVBT) produce deterministic values differing from the threshold by a given delta. Such delta is selected by the engineers to produce deterministic mutated data values that should trigger a change in software behaviour (i.e., different output) and, therefore, be detected by the test suite. 
The same holds for operators that alter data item instances that are supposed to follow constant or periodic functions (i.e., ASA, HV, SS). 
Only three operators (i.e., BF, INV, and FVOR) generate random mutated data values. 
Operators that produce random data values belonging to a valid range (i.e., INV and FVOR) are used when any data value should trigger the desired change in  behaviour while a predefined data value may lead to unexpected results; for example, when IP addresses are dynamically assigned, to cause a message loss, replacing a destination IP address with any another IP in the range is preferable to using a predefined IP value which may, for example, accidentally match the destination IP. Finally, the bit flip (BF) operator simulates the generation of noise, which, by definition, is random (i.e., BF mutates a subset of the bits selected by the engineers); however, to eliminate non-determinism, engineers can configure BF to mutate all the bits in the specified set (i.e., by setting VALUE equal to $\mathit{MAX} - \mathit{MIN}$). With non-deterministic operators, engineers can still debug executions by relying on execution traces with the mutated data values generated.
Finally, deterministic operators may prevent the identification of unforeseen critical inputs not correctly processed by the SUT; however, such an objective goes beyond the purpose of \APPR, which aims to assess test suites, not to perform robustness testing. Extensions of \APPR to achieve this goal will be the object of future work.}

Our partners confirmed the appropriateness of every operator in Table~\ref{table:operators} to  simulate plausible and critical interoperability faults in CPS software; our partners also confirmed that they could not recall other types of data modifications leading to interoperability problems.
Although other data representation types (e.g., null terminated strings) and operators (e.g., replacement of a random char in a string) might be envisioned, in this paper, we focus on operators that are critical in the CPS context, 
\REVIEW{B.9}{based on our discussions with domain experts.}
For example, CPS components are unlikely to exchange strings.

\REVIEW{C.2}{Our mutation operators can be applied to other data structures than buffers as they apply to data values. In future work, additional mutation operators can be defined to address specific aspects of other data structures, e.g., changes to structures of trees or lists.}

\subsection{Fault Modelling Methodology (Step 1)}
\label{sec:methodology}

The fault model shall enable the specification of 
all possible interoperability problems in the SUT while minimizing equivalent and redundant mutants.
Equivalent mutants have the same observable output as the original SUT. 
Instead, redundant mutants have the same observable output as other mutants.
We use the term \emph{observable output} to refer to any output that can be verified by the test suite.
The equivalent or redundant nature of a mutant depends
on the equivalence relation for observable outputs
(i.e., how to determine if two outputs are the same).
In a testing context, such equivalence relation depends on the type of testing being performed. For example, system test cases, different than unit test cases,  are unlikely to verify the values of all the state variables of the system and thus mutants that are nonequivalent for unit test suites might be considered equivalent for system test suites. 
For example, in satellite systems, the correctness of the GPS triangulation algorithm output is verified by unit test cases; system test cases, instead, verify 
if the software takes appropriate actions when the satellite is out of orbit. Consequently, slight changes in the coordinates communicated by the GPS component may not lead to any change in the observable output verified by the test suite.

\REVIEW{B.4}{\APPR assumes that the mutated data is not automatically corrected by  software 
(e.g., through cyclic redundancy check codes),
 thus leading to different execution behaviors than expected in the test cases. To target data that is automatically corrected (i.e., to assess if automated data correction is accurately verified by the SUT test suite),
it is necessary to work with test suites capable of detecting the presence of unexpected corrections (e.g., by verifying correction counters); otherwise, mutations would lead to equivalent test executions.}

\REVIEW{B.6-8}{The \APPR fault modelling methodology includes the following activities:
\begin{enumerate}
    \item Identifying, within the SUT, the messages to be mutated; each message will be targeted by a different fault model.
    \item Identifying, for each message, the data items that constitute the message and their characteristics, such as \emph{the position} of each data item in the buffer, \emph{their span}, and their \emph{representation type} (e.g., integer, double, binary).
    \item Selecting a subset of data-driven mutation operators to apply from Table~\ref{table:operators}; the selection shall be based on a set of guidelines provided below.
    \item Configuring each selected mutation operator based on software specifications, according to our guidelines.
    \item Writing a specification in tabular form (i.e., a \emph{CSV} file), where each row represents the configuration of a mutation operator for a specific data item, from a specific fault model. All the fault models of the SUT can be specified in the same specification document. An example is provided in Table~\ref{table:faultModel} and commented at the end of this Section.
\end{enumerate}}

\begin{table}[tb]
\caption{\APPR fault modelling methodology}
\label{table:method}
\scriptsize
\begin{tabular}{|
@{\hspace{1pt}}>{\raggedleft\arraybackslash}p{12mm}@{\hspace{1pt}}|
@{\hspace{1pt}}>{\raggedleft\arraybackslash}p{17mm}@{\hspace{1pt}}|
@{\hspace{1pt}}>{\raggedleft\arraybackslash}p{16mm}@{\hspace{1pt}}|
@{\hspace{1pt}}>{\raggedleft\arraybackslash}p{12mm}@{\hspace{1pt}}|
@{\hspace{1pt}}>{\raggedleft\arraybackslash}p{14mm}@{\hspace{1pt}}|
@{\hspace{1pt}}>{\raggedleft\arraybackslash}p{12mm}@{\hspace{1pt}}|
}
\hline
\textbf{Data} \textbf{nature}&\textbf{Representation} \textbf{type}&\textbf{Dependencies}&\textbf{\# of input} \textbf{partitions}&\textbf{Operators}&\textbf{Comments}\\
\hline
numerical&I, L, F, D&stateless or stateful&2&[VAT,FVAT]&Nominal below T\\
&&&&or [VBT,FVBT]&Nominal above T\\
\cline{4-6}
&&&3 or more&[VOR,FVOR]&\\
\cline{3-6}
&&stateful&&INV&For valid range\\
\cline{4-6}
&& &&[VOR,FVOR]&For out of range\\
\cline{3-6}
&&signal&&ASA, SS, HV&\\
\hline
categorical&I, H&N/A&N/A&IV&\\
\cline{2-6}
&B&N/A&N/A&BF&\\
\hline
ordinal&I, H&N/A&N/A&ASA&\\
\hline
other&B&N/A&N/A&BF&\\
\hline
\end{tabular}
\vspace{1mm}

\textbf{Legend:} N/A not applicable. I: INT, L: LONG INT, F: FLOAT, D: DOUBLE, B: BIN, H: HEX. [] complementary pair of operators.
\end{table}

We provide a set of guidelines for the \REVIEW{B.6-8}{selection and configuration of mutation operators}
that 
are summarized in Table~\ref{table:method}. 
For guidance,
we account for the nature of the data (i.e., numerical, categorical, ordinal, or binary) and their representation type.
Also, for numerical data, 
we consider 
\REVIEW{B.10}{the data dependencies, that is how data values depend on previously observed values; we identified three categories: (1) \emph{stateless} (i.e., there are no dependencies between consecutive values), (2) \emph{stateful}, when values depend on previous ones (e.g., messages sequence identifier), and (3) \emph{signal} when values are determined by a function of independent variables like time.}
Data dependencies determine the granularity of the mutation (i.e., with data dependencies, small differences shall be noticed, as explained below); for non-numerical data, since data dependencies are not observed, we do not provide mutation operators with different granularities.

For \emph{stateless numerical data}, our guidelines are driven by input space partitioning concepts~\cite{Ammann:Offutt:2008}.
Indeed, given equivalence relations among outputs, it is unlikely that every change in \emph{stateless numerical data} will result into nonequivalent mutants; however, we can partition the input domain into regions with equivalent values (partitions).
Precisely, we rely on the  
\emph{interface-based input domain modeling} approach~\cite{Ammann:Offutt:2008}:
for each data item we identify a number of input partitions (set of values or value ranges) according to the interface specifications of the interacting components.
In our methodology, the number and type of mutation operators selected for stateless numerical data depend on the number of input partitions identified.

\begin{figure}[ht]
	\centering
		\includegraphics[width=6cm]{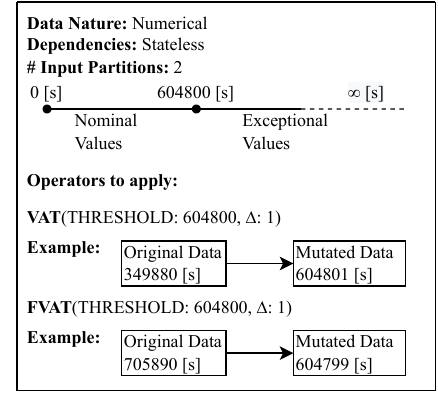}
 		\caption{Example of the \APPR methodology applied to stateless numerical data with two input partitions.}
		\label{fig:appr:example_numerical_2}
	\end{figure}

With \emph{two input partitions} (e.g., nominal and exceptional data values), engineers can rely either on the pair [VBT,FVBT] or the pair [VAT,FVAT]. An example is provided in Figure~\ref{fig:appr:example_numerical_2}; it concerns a data item with two input partitions for nominal (i.e., 0 to 604,800) and exceptional values (i.e., above 604,800). The VAT and FVAT operators are configured with the threshold parameter set to 604,800 and delta (\D) set to 1 (often, we use \D as the step granularity). At runtime, the mutant integrating the VAT operator will replace any value below 604,800 with 604,801 (i.e., $604,800 + 1$); the mutant with the FVAT operator will replace any value above 604,800 with 604,799 (i.e., $604,800 - 1$).
	
\begin{figure}[ht]
	\centering
		\includegraphics[width=6cm]{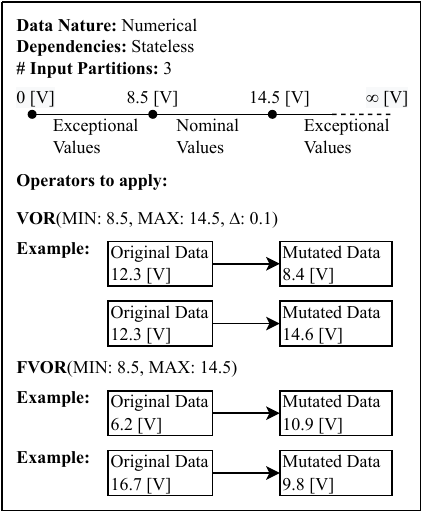}
 		\caption{Example of the \APPR methodology applied to stateless numerical data with three input partitions.}
		\label{fig:appr:example_numerical_3}
	\end{figure}

With \emph{three input partitions}, engineers must configure one VOR and one FVOR operator. Figure~\ref{fig:appr:example_numerical_3} shows the case of a data item used to transmit voltage information; it presents an input partition for nominal values (i.e., 8.5~[V] to 14.5~[V]) and two input partitions for exceptional values (i.e., below $8.5$ and above $14.5$). If a different delta (\D) is considered for the upper and lower bounds, engineers may configure two pairs [VBT,FVBT] and [VAT,FVAT], for the lower and upper bounds, respectively (see items 1 to 4 in Table~\ref{table:faultModel}, described below). 
	
In the presence of \emph{more than three} input partitions, engineers shall configure one [VOR,FVOR] pair for each extra partition above three (e.g., two pairs in the case of five partitions). The parameter \D is used to determine the partition to which the mutated data belongs. 

\begin{figure}[t]
	\centering
		\includegraphics[width=6cm]{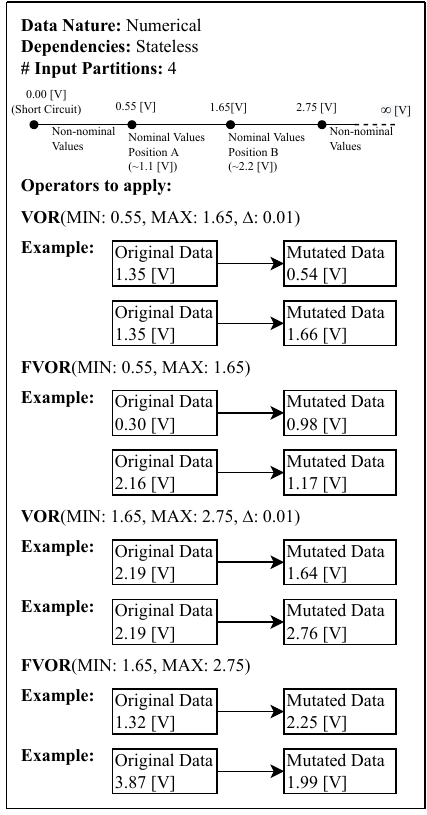}
 		\caption{Example of the \APPR methodology applied to stateless numerical data with four input partitions.}
		\label{fig:appr:example_numerical_4}
	\end{figure}

The example in Figure~\ref{fig:appr:example_numerical_4} shows four partitions for the Voltage of a \emph{TCTM} (TeleMetry and TeleCommand) switch in the \emph{ADCS} of \ESAIL. Voltage values below $0.55~[V]$ or above $2.75~[V]$ are considered non-nominal values. Among nominal values, the voltage is between $0.55~[V]$ and $1.65~[V]$ for Position A of the switch and between $1.66~[V]$ and $2.75~[V]$ for Position B.
The first mutation operator (i.e., VOR with MIN set to $0.55$, MAX set to $1.65$, and \D set to $0.01$) leads to replacing nominal values for Position A with non-nominal values below $0.55~[V]$ (i.e., $0.54$) and nominal values belonging to Position B (i.e., $1.66$).
The second mutation operator (i.e., FVOR with MIN set to $0.55$, MAX set to $1.65$, and \D set to $0.01$) leads to replacing values below $0.55~[V]$ and values belonging to Position B (i.e., $2.16$)
with random nominal values for Position A (e.g., $0.98$ and $1.17$). 
The third mutation operator (i.e., VOR with MIN set to $1.65$, MAX set to $2.75$, and \D set to $0.01$) leads to replacing nominal values for Position B with values belonging to Position A (i.e., $1.64$) and exceptional values above $2.75~[V]$ (i.e., $2.76$). 
The fourth mutation operator (i.e., FVOR with MIN set to $1.65$ and MAX set to $2.75$) leads to replacing values below $1.65~[V]$ (e.g., belonging to Position A) and non-nominal values above $2.75~[V]$ with random nominal values for Position B. 
	
\begin{figure}[t]
	\centering
		\includegraphics[width=6cm]{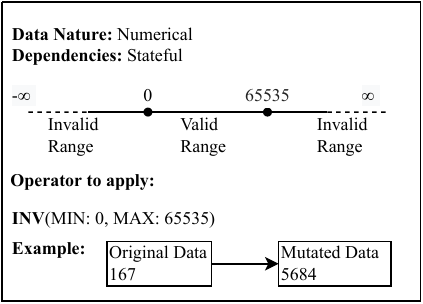}
 		\caption{Example of the \APPR methodology applied to stateful numerical data.}
		\label{fig:appr:example_stateful}
	\end{figure}

In the presence of \emph{stateful data}, replacement with random values in the valid range (i.e., the INV operator) will lead to nonequivalent mutants (e.g., because it leads to data values that are systematically different than the values expected for the current system state).
\REVIEW{B.10}{For example,  since the sequence ID of a message depends on the system state (e.g., the number of sent messages), the INV operator shall be used to replace a correct identifier with another one, leading to a failure (i.e., the SUT shall notice that the message has an invalid ID and report an output that captures such anomaly, which shall then be detected by the test suite). Figure~\ref{fig:appr:example_stateful} illustrates this case.}

When there is no guarantee that applying the INV operator makes the SUT behave differently (e.g., because the SUT extracts the max value in a sequence), the valid data range might be partitioned as for stateless data. However, to avoid redundant mutants, engineers should rely either on the INV operator or the partitioning of the valid data range. 
The effect of data outside the valid data range should instead be verified by means of the [VOR, FVOR] pair.

\begin{figure}[t]
	\centering
		\includegraphics[width=6cm]{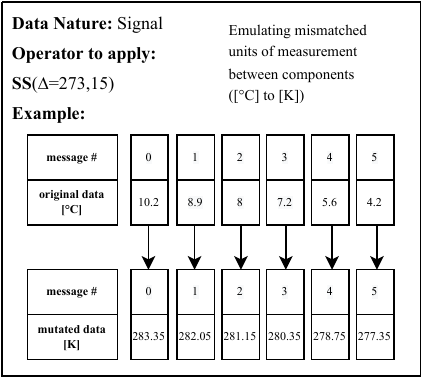}
 		\caption{Example of the \APPR methodology applied to a signal.}
		\label{fig:appr:example_signal}
	\end{figure}

\begin{figure}[t]
	\centering
		\includegraphics[width=6cm]{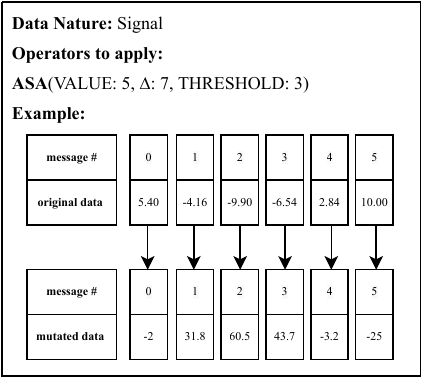}
 		\caption{Example of the ASA operator applied to a signal.}
		\label{fig:appr:example_signal_ASA}
	\end{figure}

\begin{figure}[t]
	\centering
		\includegraphics[width=6cm]{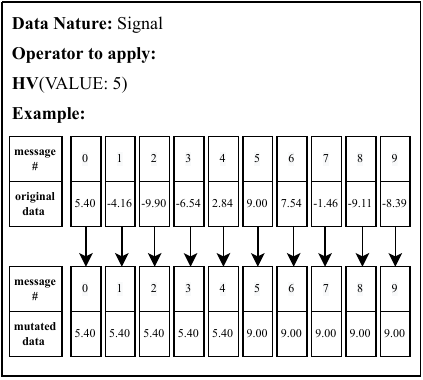}
 		\caption{Example of the HV operator applied to a signal.}
		\label{fig:appr:example_signal_HV}
	\end{figure}

For \emph{signal values}, depending on the shape of the expected signal, engineers should configure one operator among ASA, SS, and HV. The configuration of more than one of these operators may lead to redundant mutants (e.g., because each of them triggers the same warning in the SUT).
In the example of Figure~\ref{fig:appr:example_signal}, the SS operator is used to simulate a mismatch in measurement units; indeed we add 273.15 to report Kelvin degrees instead of Celsius.
Figures~\ref{fig:appr:example_signal_ASA} and \ref{fig:appr:example_signal_HV} exemplify the cases for the ASA and HV operators, respectively. In Figure~\ref{fig:appr:example_signal_ASA}, values above (below) 3 are incremented (decremented) according to the formula in Table~\ref{table:operators}. Figure~\ref{fig:appr:example_signal_HV} shows that the same value is repeated five times, then the newly observed value (i.e., 9.00) is repeated another five times, and so on.

\begin{figure}[t]
	\centering
		\includegraphics[width=6cm]{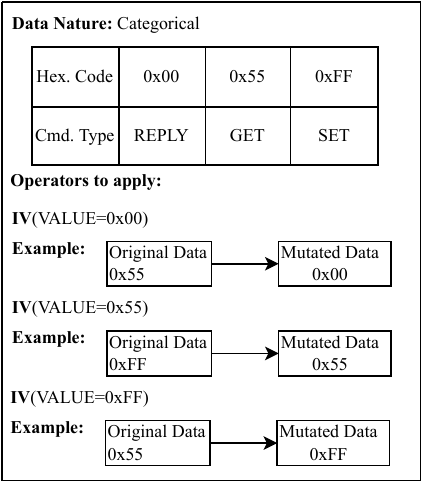}
 		\caption{Example of the \APPR methodology applied to categorical data in hexadecimal format.}
		\label{fig:appr:example_categorical}
	\end{figure}

With \emph{categorical data} represented using \emph{integers and hexadecimals}, engineers must configure one IV operator for each possible value; indeed, a change in the observed category shall trigger a different behaviour in the SUT.
Figure~\ref{fig:appr:example_categorical} provides an example where the IV operator is used to alter the command to be executed by the component receiving the message; we configure three IV operators because the system under test includes three commands.

\begin{figure}[t]
	\centering
		\includegraphics[width=6cm]{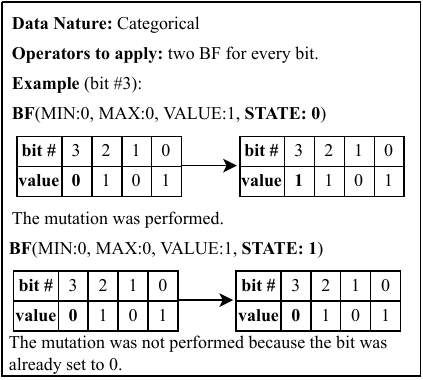}
 		\caption{Example of the \APPR methodology applied to categorical data in binary format.}
		\label{fig:appr:example_categorical_binary}
	\end{figure}

With categorical data in \emph{binary form}, each bit indicates a specific class (e.g., the unit in error for the DataItem2 in the IFStatus message of Figure~\ref{fig:appr:bufferStructure}).  
To verify that the test suite can detect any possible category change, engineers must configure two BF operators for every bit (both MIN and MAX must coincide with the bit position), one operator must flip a bit when it is set (i.e., $\mathit{STATE}=1$), and the other one when it is unset (i.e., $\mathit{STATE}=0$); an example appears in Figure~\ref{fig:appr:example_categorical_binary}.  
	
\begin{figure}[t]
	\centering
		\includegraphics[width=6cm]{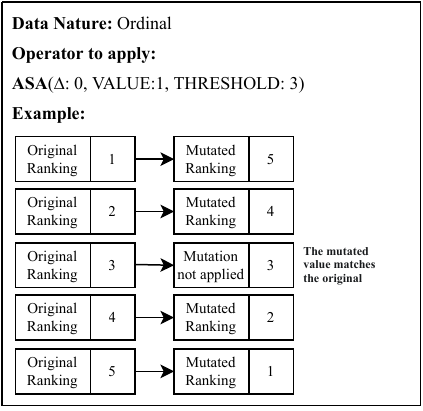}
 		\caption{Example of the \APPR methodology applied to ordinal data.}
		\label{fig:appr:example_asa_ordinal}
	\end{figure}

For \emph{ordinal data}, which is represented by means of either integers or hexadecimals, we suggest to apply the ASA operator with \emph{T} being set to the middle point of the ordinal scale and \emph{VALUE} set to the step distance between consecutive data (usually $1$) as exemplified in Figure~\ref{fig:appr:example_asa_ordinal}.

\begin{figure}[t]
	\centering
		\includegraphics[width=6cm]{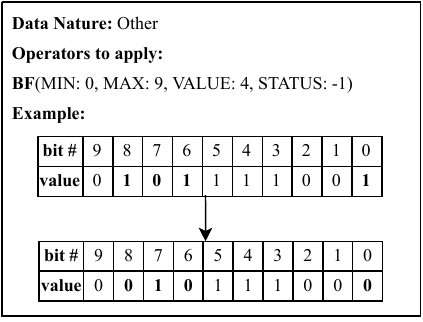}
 		\caption{Example of the \APPR methodology applied to generic data in binary form.}
		\label{fig:appr:example_other}
	\end{figure}

For data in \emph{binary form} (e.g., pictures), engineers must configure a BF operator to flip a number of bits  that is sufficient to alter the semantics of the data (e.g., introduce sufficient noise in images). Figure~\ref{fig:appr:example_other} shows a BF operator that, at runtime, flips four randomly selected bits.

\begin{table}[tb]
\caption{Portion of the fault model specification for \ESAIL}
\label{table:faultModel}
\tiny
\begin{tabular}{|
@{\hspace{1pt}}>{\raggedleft\arraybackslash}p{2.1mm}@{\hspace{1pt}}|
@{\hspace{1pt}}>{\raggedleft\arraybackslash}p{8.5mm}@{\hspace{1pt}}|
@{\hspace{1pt}}>{\raggedleft\arraybackslash}p{8mm}@{\hspace{1pt}}|
@{\hspace{1pt}}>{\raggedleft\arraybackslash}p{5mm}@{\hspace{1pt}}|
@{\hspace{1pt}}>{\raggedleft\arraybackslash}p{9mm}@{\hspace{1pt}}|
@{\hspace{1pt}}>{\raggedleft\arraybackslash}p{5mm}@{\hspace{1pt}}|
@{\hspace{1pt}}>{\raggedleft\arraybackslash}p{4mm}@{\hspace{1pt}}|
@{\hspace{1pt}}>{\raggedleft\arraybackslash}p{6mm}@{\hspace{1pt}}|
@{\hspace{1pt}}>{\raggedleft\arraybackslash}p{6mm}@{\hspace{1pt}}|
@{\hspace{1pt}}>{\raggedleft\arraybackslash}p{7mm}@{\hspace{1pt}}|
@{\hspace{1pt}}>{\raggedleft\arraybackslash}p{7mm}@{\hspace{1pt}}|
@{\hspace{1pt}}>{\raggedleft\arraybackslash}p{7mm}@{\hspace{1pt}}|
}
\hline
\textbf{\#}&\textbf{Fault Model}&\textbf{Position}&\textbf{Span}&\textbf{Type}&\textbf{Op}&\textbf{MIN}&\textbf{MAX}&\textbf{T}&\textbf{DELTA}&\textbf{STATE}&\textbf{VALUE}\\
\hline
1&IfHK&0&2&DOUBLE&VAT&-&-&33.53&0.01&-&-\\
2&IfHK&0&2&DOUBLE&FVAT&-&-&33.53&0.01&-&-\\
3&IfHK&0&2&DOUBLE&VBT&-&-&24&1&-&-\\
4&IfHK&0&2&DOUBLE&FVBT&-&-&24&1&-&-\\
5&IfHK&10&2&DOUBLE&VAT&-&-&6&1&-&-\\
6&IfHK&12&2&DOUBLE&VOR&-20&50&-&1&-&-\\
7&IfHK&14&2&DOUBLE&VOR&-20&50&-&1&-&-\\
8&IfStatus&0&1&BIN&BF&3&3&-&-&0&1\\
9&IfStatus&0&1&BIN&BF&3&3&-&-&1&1\\
10&IfStatus&0&1&BIN&BF&4&4&-&-&0&1\\
11&IfStatus&0&1&BIN&BF&4&4&-&-&1&1\\
12&IfStatus&0&1&BIN&BF&5&5&-&-&0&1\\
13&IfStatus&0&1&BIN&BF&5&5&-&-&1&1\\
14&IfStatus&1&1&BIN&BF&0&0&-&-&0&1\\
15&IfStatus&1&1&BIN&BF&0&0&-&-&1&1\\
16&IfStatus&1&1&BIN&BF&1&1&-&-&0&1\\
17&IfStatus&1&1&BIN&BF&1&1&-&-&1&1\\
18&IfStatus&1&1&BIN&BF&2&2&-&-&0&1\\
19&IfStatus&1&1&BIN&BF&2&2&-&-&1&1\\
20&IfStatus&1&1&BIN&BF&3&3&-&-&0&1\\
21&IfStatus&1&1&BIN&BF&3&3&-&-&1&1\\
22&IfStatus&1&1&BIN&BF&4&4&-&-&0&1\\
23&IfStatus&1&1&BIN&BF&4&4&-&-&1&1\\
\hline
\end{tabular}\\
Note: a "-" is used for parameters not required to configure a mutation operator.
\end{table}

Table~\ref{table:faultModel} provides a specification in tabular form (i.e, the format processed by \APPR) of two fault models configured for the IfKH (i.e., Interface House Keeping) and IfStatus (i.e, Interface Status) messages. In the fault models, each row captures the configuration of a mutation operator for a specific data item. For example, row number 5 indicates that \APPR interprets as double the data inside the two buffer units starting at position 10 (units 10 and 11) and applies the VAT operator. Rows 1 and 3 show that, for a same numerical data item (i.e., the one covering units 0 and 1), we can apply both the VAT and VBT operators, using a different delta for each. 
Rows 2 and 4 show the FVAT and FVBT operators complementing the VAT and VBT operators in rows 1 and 3. They simulate the case in which data for the nominal cases is observed instead of data for exceptional cases, as visible in Table~\ref{table:operators}.
Rows 8 to 23 show that different bits of a same data item can be targeted by different BF operators. %
\UPDATED{Rows 8 to 13 concern binary categorical data with two categories each, thus we configured two BF each}. 
Rows 14 to 23 concern binary categorical data with five categories; consequently, they present ten BF operators configured for the five categories.

\subsection{Automated Generation of Mutation API (Step 2) and Probe Insertion (Step 3)}
\label{sec:generateAPI}

\begin{figure}[tb]
\includegraphics[width=7cm]{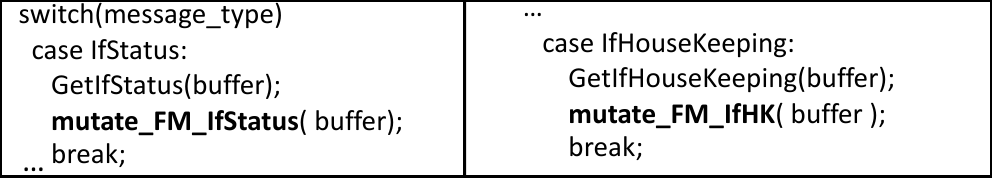}
\caption{Example of \APPR mutation probes (in bold).}
\label{fig:appr:ProbesExample}
\end{figure}

\APPR automatically generates a \emph{mutation API} to perform mutations at runtime. The API implements a set of functions (called \emph{mutate\_FM\_\textless name\textgreater}) that mutate a data buffer according to the given fault model. 
These functions select the data item to mutate and the mutation procedure to apply based on the mutant under test (see Section~\ref{sec:mutantsGeneration}). 

The \APPR mutation API works with C/C++ code; however  it may be extended to deal with other programming languages.
Since it is not possible to automatically determine which data buffer to mutate, \APPR requires engineers to modify the source code of the CPS under test by introducing a mutation probe which consists of an invocation of the \APPR function that mutates the data buffer according to a specific fault model. 

\REVIEW{B.5}{Mutation probes should be inserted into the functions that handle components' communication within either the SUT or the simulator. In our experiments, we inserted them into functions that either receive or produce data (see Section~\ref{sec:subjects}). We expect engineers to insert probes into functions that are easier to modify (e.g., requiring fewer probes, with more lenient time requirements).} 

\REVIEW{C.8}{We insert probes into the source code of the software under test (or the simulator used to drive testing) instead of the transmission layer (e.g., OS network functions) because data mutation is driven by the data semantics, which varies according to message type (e.g., the thresholds for VBT). Since the transmission layer does not distinguish low-level messages based on their content, we insert probes into code locations (i.e., SUT functions) where the message type is known. Probes inserted into the transmission layer would need to process the message content to match it to message types (e.g., through IDs); such a solution would introduce runtime overhead and further modeling effort.}

Note that the effort required to insert probes is limited; indeed, the exchange of data between components is usually managed in a single location (e.g, the function that serializes the data buffer on the network) and thus it is usually sufficient to introduce one function call for each message type to mutate.

Figure~\ref{fig:appr:ProbesExample} shows how the implementation of \ESAIL has been modified to add the mutation probes. 
The SVF function was modified to handle the message requests sent to the ADCS by inserting one mutation probe for each message type to mutate, e.g., IfStatus and IfHouseKeeping in~Figure~\ref{fig:appr:ProbesExample}. 
Function \emph{mutate\_FM\_IfStatus} is part of the generated mutation API; it loads the fault model \emph{IfStatus} into memory (\UPDATED{our API relies on a tree data structure}) and then invokes the function \emph{mutate}. The function \emph{mutate} performs data-driven mutation according to the provided fault model; the implementation of \emph{mutate} is part of the \APPR toolset.

The behavior of function \emph{mutate} depends on the value of a unique identifier (i.e., the \emph{MutantID}) associated at compile time to the mutant; the \emph{Mutant ID} univocally identifies the performed mutation operation (each mutant executes one mutation operation, see Section~\ref{sec:mutantsGeneration}).
At a high level, \emph{mutate} performs four activities. First, it checks if the mutation should be performed (i.e., if the data buffer is targeted by the mutation operation identified with the \emph{Mutant ID}). Second, it casts the data item instance targeted by the mutant to a support variable of the type specified in the fault model. Third, it mutates the data stored in the support variable; for each mutation operator, we have implemented a distinct set of instructions for each data representation type. Fourth, before terminating, the function \emph{mutate} writes the mutated data back to the data buffer.

\subsection{Automated Generation of Mutants (Step 4)}
\label{sec:mutantsGeneration}

Consistent with code-driven mutation analysis, \APPR generates one mutant for each mutation procedure of the mutation operators configured in the fault model. Each mutant performs exactly one \emph{data mutation operation} (i.e., a data mutation procedure configured for a specific data item). For example, the specification in row 6 of Table~\ref{table:faultModel} makes \APPR generate two mutants: each mutant modifies the value of the data item starting at position 12 but one mutant replaces the current value with the value 51 (i.e., $50+1$) while the other replaces the current value with the value $-21$ (i.e., $-20 -1$).

The mutant generation is invisible to the end-user who does not need to modify the source code further; indeed, we rely on a C macro to specify, at compile time, which mutation operation must be performed by every mutant. Mutants are generated by compiling the \UPDATED{SUT} multiple times, once for each mutation operation. At runtime, the mutate function executes only the mutation operation selected for the mutant under test.

\subsection{Mutants Execution (Step 5)}
\label{sec:mutantsExecution}

As for code-driven mutation analysis, the test suite under analysis is executed iteratively with every data-driven mutant. 
At runtime, all the data items targeted by a mutant are mutated whenever the mutation preconditions hold (e.g., the STATE of the BF operator); we leave the mutation of a sampled subset of \UPDATED{data item instances to future work~\cite{zhang2013operator,gopinath2015hard}.}
To speed up the mutation analysis process, the test suite under analysis is first executed with a special mutant that, instead of mutating data items, keeps trace of the fault models loaded by each test case; in other words, it traces what are the data types covered by each test case. The collected information enables the execution, for every mutant, of the subset of test cases that cover the 
message type
targeted by the mutant, thus speeding up mutation analysis.

\begin{figure}[ht]
	\centering
		\includegraphics[width=8cm]{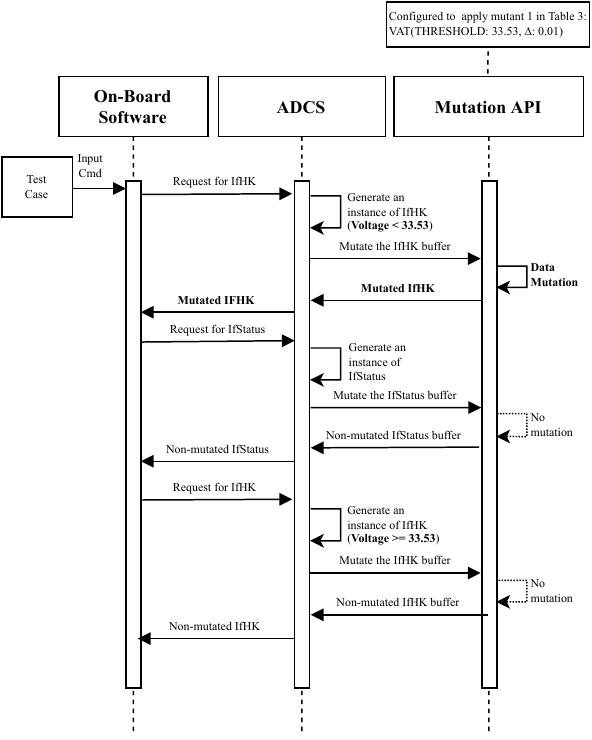}
 		\caption{UML sequence diagram for the execution of an \ESAIL test case.}
		\label{fig:appr:esail_sequence}
	\end{figure}

The UML sequence diagram in Figure~\ref{fig:appr:esail_sequence} exemplifies the execution of a test case for \ESAIL with a mutant generated by the operator configured in the first row of Table~\ref{table:faultModel}, targeting the \emph{IfHK} message. 
During the test case execution, there are three interactions between the on-board software and the ADCS. The first interaction is a request for \emph{IfHK}, which makes the ADCS generate, in a buffer, an \emph{IfHk} message with a voltage value below 33.53 (DataItem 1 in Figure~\ref{fig:appr:bufferStructure}). 
During the execution, the buffer is mutated by the \APPR probe through a call to the \emph{Mutation API}.
The second interaction is a request for an \emph{IfStatus} message; in this case the message is not mutated because, in the current mutant, the API is not configured for targeting \emph{IfStatus} messages.
The third interaction is a request for \emph{IfHK}, which is the mutant target. However, in this case, the data is not mutated because the voltage value is above 33.53 (i.e., the threshold for the VAT operator).

\subsection{Mutation Analysis Results (Step 6)}
\label{sec:mutationAnalysisResults}

Inspired by work on abstract mutation analysis~\cite{Offutt2006}, we have defined three metrics to evaluate test suites with data-driven mutation analysis: fault model coverage, mutation operation coverage, and covered mutation score. 
These metrics measure the frequency of the following scenarios: (case 1) the message type targeted by a mutant is never exercised, (case 2) the message type is covered by the test suite but it is not possible to perform some of the mutation operations (e.g., because the test suite does not exercise out-of-range cases), (case 3) the mutation is performed but the test suite does not fail.

\emph{Fault model coverage (FMC)} is the percentage of fault models covered by the test suite. Since we define a fault model for every 
message type exchanged by two components,
it provides information about the extent to which the message types actually exchanged by the SUT are exercised and verified by the test suites. 
\CHANGED{Since different component functionalities often require different message types, low fault model coverage may indicate that only a small portion of the integrated functionalities have been tested.}

\emph{Mutation operation coverage (MOC)} is the percentage of data items that have been mutated at least once, considering only those that belong to the data buffers covered by the test suite. It provides information about the input partitions covered for each data item; for example, the FVOR operator leads to two mutation operations, which are applied only if the observed value is outside range. Otherwise the two mutation operations will not be covered, thus enabling the engineer to identify such shortcoming in the test suite.

\TSEmin{2.2}{The \emph{covered mutation score (CMS)}} is the percentage of mutants killed by the test suite \UPDATED{(i.e., leading to at least one test case failure)} among the mutants that target a fault model and for which at least one mutation operation was successfully exercised (i.e., covered). It provides information about the quality of test oracles; indeed, a mutant that performs a mutation operation and is not killed (i.e., is \emph{live}) indicates that the test suite cannot detect the effect of the mutation (e.g., the presence of warnings in logs).
\CHANGED{Also, a low CMS may indicate missing test input sequences. Indeed, live mutants may be due to either software faults (e.g., the SUT does not provide the correct output for the mutated data item instance) or the software not being in the required state (e.g., input partitions for data items are covered when the software is paused); in such cases, with appropriate input sequences, the  test suite would have discovered the fault or brought the SUT into the required state. Both poor oracles and lack of inputs indicate flaws in the test case definition process (e.g., the stateful nature of the software was ignored).}

\REVIEW{A.5}{Different from the ones produced by code-driven mutation analysis, these three metrics enable engineers to distinguish among possible test suite shortcomings. A test suite shortcoming means that the test suite is not exercising a specific input or message or that oracles are not being (fully) checked. A test suite shortcoming prevents the detection of observable failures caused by data---exchanged between SUT components---not being equivalent to the data assumed by test cases. The categories of test suite shortcomings include (1) untested message types, (2) uncovered input partitions, (3) poor oracle quality, and (4) lack of test inputs.}
\REVIEW{B.2}{\emph{Fault model coverage} spots untested message types. \emph{Mutation operation coverage} reports uncovered partitions and is computed after excluding mutants belonging to untested message types. The \emph{covered mutation score} is computed after excluding mutants that implement a mutation operation that was not applied; for this reason, a low CMS indicates poor oracle quality and lack of test inputs bringing the system into a specific state.}

\section{Empirical Evaluation}
\label{sec:empirical}

We address the following research questions:

\emph{RQ1. What are the types of test suite shortcomings identified by \APPR?}
    We aim to assess the effectiveness of \APPR in identifying various test suite shortcomings, as described in Section~\ref{sec:mutationAnalysisResults}. In other words, we want to know if mutation analysis based on \APPR can provide clear guidance in terms of what to improve in a test suite.

\emph{RQ2. 
What is the impact of equivalent and redundant data-driven mutants on the mutation analysis process?}
    In general, mutation analysis may lead to the generation of equivalent and redundant mutants.  
    In the specific context of \APPR, we analyze the extent of their impact on mutation scores.

\emph{RQ3.  Is data-driven mutation feasible?}
    To assess its feasibility in practice, we evaluate the cost of setting up data-driven mutation analysis (i.e., defining fault models and instrumenting the CPS with probes), \UPDATED{the duration of the mutation analysis process, and the} runtime overhead introduced during test case execution.    

\subsection{Subjects of the study}
\label{sec:subjects}

To assess our research questions, we considered 
CPS components used in cubesat constellations and in \SAIL, which is a micro-satellite~\cite{satSurvey} launched into space \LAUNCH{}. 
More precisely, we consider \PARAM, which is a client-server component to manage configuration parameters in cubesats.
Also, we examine three \SAIL software sub-systems (1) the Attitude Determination And Control System (\ADCS), the Global Positioning System (\GPS), and the Payload Data Handling Unit (\PDHU). \UPDATED{These are representative examples of CPS control and utility software, as well as sensor and actuator drivers.}

We rely on \APPR to evaluate the \PARAM integration test suite by  mutating the data exchanged between the client and server components of \PARAM.
Similarly, \APPR is used to evaluate how well the 
\SAIL test suite covers interoperability problems affecting the integration between the control software of \SAIL (hereafter, CSW) and the \ADCS, \PDHU, and \GPS components. We thus
mutate the data exchanged between \SAIL CSW and these three components.
Since each of these sub-systems have a  different purpose (i.e., their data is processed by distinct CSW functions and affect distinct \SAIL features) we treat them as distinct case study subjects although they are tested using the same test suite. We focus on the \SAIL test suite that makes use of an SVF to simulate the \ADCS, \PDHU, and \GPS components. 
The main reason is that these three components can only be executed on the target hardware and thus most of the scenarios involving them are tested in a simulated environment first.
\CHANGED{We do not mutate messages or data items that are tested only with HIL.}

In the case of \PARAM, we inject mutation probes into the \PARAM server to mutate both received and generated messages. For \SAIL, we insert mutation probes into the SVF that mutate the messages it generates; we avoid mutating the messages received by the SVF because such mutations may lead to input data it does not support. 
\ESAIL features 74 kLoC and its SVF 65 kLoC. 
The \ESAIL test suite includes 384 test cases, takes approximately 10 hours to execute, and relies on three simulated
\SVF sub-systems (i.e., \ADCS, \GPS, and \PDHU).
Instead, \PARAM contains 3 kLoC and is tested through an integration test suite which is composed by 170 test cases. 
The \PARAM integration test suite takes approximately 1 minute to execute.
We have executed all the test suites with every mutant without imposing time limits\TSEmin{3.9}{, except for test timeouts: a test case is considered to be failing if its execution takes more than three times the time required with the original SUT.}
By considering both a quick integration test suite and an extensive system test suite, we aim to cover the diversity of scenarios in which our approach can be applied.

\subsection{Experimental Setup}

With the support of our industry partners, we relied on  
 the systems' specification documents 
 to define the fault models for each subject.

\begin{table}[tb]
\caption{Fault models and mutation operators.}
\label{table:summary} 
\scriptsize
\centering
\begin{tabular}{|
@{\hspace{1pt}}p{16mm}@{\hspace{0pt}}|
@{\hspace{0pt}}>{\raggedleft\arraybackslash}p{17mm}@{\hspace{1pt}}|
@{\hspace{0pt}}>{\raggedleft\arraybackslash}p{25mm}@{\hspace{1pt}}|
@{\hspace{0pt}}>{\raggedleft\arraybackslash}p{25mm}@{\hspace{1pt}}|
p{4mm}|}
\hline
\textbf{Subject}&\textbf{Fault Models}&\textbf{Configured Operators}&\textbf{Mutation Operations}\\
\hline
\ADCS& 10 & 142 & 172 \\
\GPS& 1 & 23 & 23 \\
\PDHU& 3 & 29 & 29 \\
\PARAM& 6 & \NEWRES{44} & \NEWRES{44} \\
\hline
\end{tabular}
\end{table}

\UPDATED{Table~\ref{table:summary} provides information about the fault models.}
The fault models (FMs) for the \ADCS include multiple configurations (\emph{Configured operators}) of eight mutation operators: BF, VAT, VBT, VOR, IV, FVOR, FVBT, and FVAT.
The \PDHU fault models include four operators: BF, IV, VAT and FVAT.
Even though the \GPS fault model concerns only one data type, it makes use of six operators: ASA, HV, IV, SS, VAT, and FVAT.
For \PARAM, we relied on the operators BF, HV, IV, SS, VAT, and FVAT. \NEWRES{They have led to 172 mutation operations for \ADCS, 23 for \GPS, 29 for \PDHU, and 44 for \PARAM; the number of configured operators and mutation operations match except when we rely on VOR and FVOR.}

\TSEmin{3.10}{Except for INV,} all the mutation operators provided by \APPR have been used in at least one fault model, which shows their usefulness. \TSEmin{3.10}{The INV operator replaces the observed value with another one in the valid domain; to demonstrate its usefulness, we used it to mutate the destination port and address of an opensource network library produced by \GomSpace and tested it in our tutorial on \APPR~\cite{DAMAT:Tutorial:Web}, which we did not include in our experiments.}

We performed our experiments using an HPC cluster with Intel Xeon E5-2680 v4 (2.4 GHz) nodes.

\subsection{RQ1 - Approach effectiveness}

We analyzed the extent to which \APPR helps identify limitations in test suites.
For each subject, we inspected uncovered fault models, uncovered mutation operations, and live mutants. We then analyzed how they could potentially be explained by the types of shortcomings introduced in Section~\ref{sec:mutationAnalysisResults}:
untested message types (UMT), uncovered input partitions (UIP), poor oracle quality (POQ), and lack of test inputs (LTI).
To achieve the above, we proceeded as follows.
For each uncovered fault model, we discussed with developers if the functionality triggering the exchange of the targeted message was tested by the test suite.
For uncovered mutation operations, we discussed with engineers if they match an uncovered input partition.
For live mutants, we determined if they could be killed by improving test oracles (see how equivalent mutants are detected for RQ2).

To address RQ1, based on the above analysis, we discuss below how our metrics (i.e., \emph{fault model coverage - FMC}, \emph{mutation operation coverage - MOC}, and \emph{covered mutation score - CMS}) relate to the predefined shortcoming categories \UPDATED{(e.g., a low CMS may indicate missing test oracles)}.
\CHANGED{Further, to understand how variations in test effectiveness could be explained, we investigate how our metrics relate to the number of functionalities under test (i.e., the number of fault models - $\mathit{FM}$), the number of mutation operations ($MO$), and the number of covered mutation operations ($CMO$), respectively. 
To get an idea of observable trends, we compute the Spearman's correlation coefficients between them, hereafter denoted $\rho_{FM}$, $\rho_{MO}$, $\rho_{CMO}$.}

\subsubsection*{Results}

\begin{table}[tb]
\caption{Mutation Analysis Results.}
\label{table:mutationresults} 
\scriptsize
\begin{tabular}{|
@{\hspace{0pt}}>{\raggedleft\arraybackslash}p{16mm}@{\hspace{1pt}}|
@{\hspace{0pt}}>{\raggedleft\arraybackslash}p{7mm}@{\hspace{1pt}}|
@{\hspace{0pt}}>{\raggedleft\arraybackslash}p{9mm}@{\hspace{1pt}}|
@{\hspace{0pt}}>{\raggedleft\arraybackslash}p{14mm}@{\hspace{1pt}}|
@{\hspace{0pt}}>{\raggedleft\arraybackslash}p{9mm}@{\hspace{1pt}}|
@{\hspace{0pt}}>{\raggedleft\arraybackslash}p{8.5mm}@{\hspace{1pt}}|
@{\hspace{0pt}}>{\raggedleft\arraybackslash}p{7.5mm}@{\hspace{1pt}}|
@{\hspace{0pt}}>{\raggedleft\arraybackslash}p{5.5mm}@{\hspace{1pt}}|
@{\hspace{0pt}}>{\raggedleft\arraybackslash}p{9mm}@{\hspace{1pt}}|
}
\hline
\textbf{Subject} & 
\textbf{\# FMs} & 
\textbf{FMC} & 
\textbf{\#MOs-CFM} & 
\textbf{\#CMOs} & 
\textbf{MOC}  
&\textbf{Killed}&\textbf{Live}&\textbf{CMS}
\\
\hline

\ADCS &10 &90.00\%   & 135 & 100 & 74.00\%   &    45&55&45.00\%\\
\GPS &1 &100.00\%    &  23  &  22 & 95.65\%    &      21&1&95.45\%\\
\PDHU &3 &100.00\%  &   29 & 24 & 82.76\%   &     24&0&100.00\%\\
\PARAM &6 &100.00\%  &   \NEWRES{44} & \NEWRES{41} & \NEWRES{93.20}\%  &        \NEWRES{37}&\NEWRES{4}&\NEWRES{90.24}\%\\

\hline

\end{tabular}
\vspace{1mm}

CMO=Covered Mutation Operation, MOs-CFM=Mutation Operations in covered FMs.
\end{table}

\begin{table}[tb]
\caption{Shortcomings of CPSs test suites.}
\label{table:shortcomings} 
\scriptsize
\begin{tabular}{|
@{\hspace{1pt}}p{9mm}
@{\hspace{1pt}}|
@{\hspace{2pt}}>{\raggedleft\arraybackslash}p{3mm}@{\hspace{4pt}}|
@{\hspace{2pt}}>{\raggedleft\arraybackslash}p{4mm}@{\hspace{6pt}}|
@{\hspace{3pt}}>{\raggedleft\arraybackslash}p{2mm}@{\hspace{0pt}}|
>{\raggedleft\arraybackslash}p{3mm}@{\hspace{4pt}}|
@{\hspace{2pt}}>{\raggedleft\arraybackslash}p{4mm}@{\hspace{4pt}}|
@{\hspace{-1pt}}>{\raggedleft\arraybackslash}p{5mm}@{\hspace{2pt}}|
>{\raggedleft\arraybackslash}p{4mm}@{\hspace{4pt}}|
@{\hspace{2pt}}>{\raggedleft\arraybackslash}p{4mm}@{\hspace{4pt}}|
@{\hspace{2pt}}>{\raggedleft\arraybackslash}p{4mm}@{\hspace{2pt}}|
>{\raggedleft\arraybackslash}p{4mm}@{\hspace{4pt}}|
@{\hspace{2pt}}>{\raggedleft\arraybackslash}p{4mm}@{\hspace{4pt}}|
@{\hspace{2pt}}>{\raggedleft\arraybackslash}p{4mm}@{\hspace{2pt}}|
p{3mm}|}
\hline
\textbf{Short-}      & \multicolumn{3}{c|}{\textbf{\ADCS}} & \multicolumn{3}{c|}{\textbf{\GPS}} & \multicolumn{3}{c|}{\textbf{\PDHU}} & \multicolumn{3}{c|}{\textbf{\PARAM}} \\
\textbf{coming} & \textbf{UF}&\textbf{UM} &\textbf{LM} & \textbf{UF}&\textbf{UM} &\textbf{LM} & \textbf{UF}&\textbf{UM} &\textbf{LM} & \textbf{UF}&\textbf{UM} &\textbf{LM}\\
\hline 
UMT                              &1&-&-&-&-&-&-&-&-&-&-&-\\
UIP     &-&35&-&-&1&-&-&5&-&-&7&-\\
POQ         &-&-&55&-&-&1&-&-&-&-&-&\NEWRES{4}\\
LTI         &-&-&-&-&-&-&-&-&-&-&-&-\\
\hline
\textbf{Total} &1&35&55&-&1&1&-&5&-&-&7&\NEWRES{4}\\
\hline
\end{tabular}
\vspace{1mm}

UF=Uncovered Fault model, UM=Uncovered Mutation operation, LM=Live mutant.
\end{table}

Table~\ref{table:mutationresults} reports the mutation analysis results according to the metrics introduced in  Section~\ref{sec:mutationAnalysisResults}.
In Table~\ref{table:shortcomings}, we report how uncovered fault models, uncovered mutation operations, and live mutants are distributed with respect to the different shortcomings we noticed on each subject.

Concerning \emph{fault model coverage}, \ADCS reached a coverage of 90.00\%, while \GPS, \PDHU, and \PARAM all achieved 100\%.  
As expected, the much higher number of messages to test for \ADCS leads to incomplete testing. 

\ADCS reached 74\% \emph{mutation operation coverage}. \GPS, \PDHU, and \PARAM  achieved even higher coverage with 95.65\%, 82.76\%, \NEWRES{and 93.20\%},  respectively. Since 
$\rho_{MO}$ = -0.8, results suggest that lower mutation operation coverage is more likely when systems are more complex (i.e., there are many mutation operations, whose numbers depend on the number of input partitions). 

Regarding \emph{covered mutation scores}, we report 45.00\% for \ADCS,  95.45\% for \GPS,  and 100.00\% for \PDHU. These results indicate a varying performance of the \SVF test suite across sub-systems.
\NEWRES{\PARAM obtained a CMS of 90.24\%.}
Given that 
\NEWRES{$\rho_{CMO}$ = -0.8}, 
we conclude that the mutation score tends to be lower for complex systems with a large number of covered mutation operations \footnote{Please note that the covered mutation score is computed considering only the mutation operations that are exercised (i.e., covered).}; \TSEmin{3.11}{indeed, a large number of mutation operations derive from data items and input partitions in the data model that are difficult to exercise when test cases are manually defined.}

Table~\ref{table:shortcomings} provides the shortcomings identified for all our subjects. 
Our analysis confirms that (1) uncovered fault models (i.e., low \emph{FMC}) indicate lack of coverage for certain message types (\emph{UMT}) and, in turn, the lack of coverage of a specific functionality (i.e., setting the pulse-width modulation in \ADCS); (2) uncovered mutation operations (i.e., low \emph{MOC}) highlight the lack of testing of certain input partitions (\emph{UIP}); (3) live mutants (i.e., low \emph{MS}) suggest poor oracle quality (\emph{POQ}). In our case study systems the presence of live mutants was not explained by the lack of test inputs in the original test suite. Moreover, we have not uncovered latent faults, which is unsurprising given that all these systems went through all testing stages, including HIL, and are on orbit. 
\TSEmin{2.4}{If applied at earlier stages of development, \APPR could have supported the early improvement of test suites, thus likely preventing the discovery of faults at late development stages (e.g., HIL testing). Our future work includes evaluating the benefits of \APPR at earlier stages in  industrial projects.}

\subsection{RQ2 - Equivalent and redundant mutants}

As they potentially have significant impact on the applicability of any mutation analysis approach, we assess the impact of equivalent and redundant mutants generated by \APPR.

We determined if a live mutant is nonequivalent by verifying, with the support of our industry partners, 
if there existed a test case that, after performing the mutation operation, would generate one observable output (e.g., log entry, state variable, or data sent in response to test inputs) that differs from the one generated by the original program. Otherwise a mutant was considered equivalent.

According to related work, two mutants should be considered redundant if they produce the same observable output for every possible input~\cite{Shin:TSE:DCriterion:2018}.
Since, with large CPSs, it is not possible to automatically determine if such a condition holds (e.g., differential symbolic execution may not scale and is hardly applicable when components communicate through a network), we rely on manual inspection. To make such an analysis feasible, we first need to select a subset of mutant pairs that appear to be redundant (e.g., mutants that produce the same output for every executed test case). 
However, the size of the CPSs under analysis prevents the collection of all the observable outputs produced by the system. 
We thus select as seemingly redundant all the pairs of killed mutants that (1) are exercised by the same test cases and (2) present the same failing assertions for every test case. We then manually inspect the test cases to determine if an additional assertion or a different test input might lead to different results for the two mutants, thus determining if the two mutants are actually redundant. Similar to related work, we exclude live mutants from this analysis~\cite{papadakis2016threats}.

\subsubsection*{Results.} 
All live mutants (i.e., 55 mutants for \ADCS, 1 mutant for \GPS, and 45 mutants for \PARAM) generate outputs that differ from the original CPS and, therefore, we did not detect any equivalent mutant. Though it needed to be confirmed, such result was expected since our methodology (Section~\ref{sec:methodology}), if correctly applied, suggests, for every data item, a set of mutation operators that, by construction, should not lead to mutated data that is equivalent to the original data.
Live mutants can be killed by introducing oracles that (1) verify additional entries in the log files (39 instances for \ADCS, 1 instance for \GPS), (2) verify additional observable state variables (14 instances for \ADCS, 45 instances for \PARAM), and (3) verify not only the presence of error messages but also their content (2 instances for \ADCS).

We did not find redundant mutants either, which was expected since (1) mutations concerning different data items, by definition, are expected to lead to different outputs, (2) the set of operators applied to a same data item, if selected according to our methodology, cannot lead to mutated data that is redundant.
\REVIEW{B.13}{When analysing pairs of \emph{seemingly redundant mutants}, we identified five reasons why these mutants were actually not redundant}: (1) the test case does not distinguish failures across data items (e.g., temperature values collected by different sensors), 
(2) the test case does not distinguish errors across different messages (e.g., in \ADCS, the IfHK message reporting a broken sensor or the message sent by a sensor reporting malfunction), (3) the test case does not distinguish between errors in nominal and non-nominal data (e.g., it does not distinguish between VOR and FVOR), (4) the test case does not distinguish between upper and lower bounds (e.g., the mutants for VOR lead to the same assertion failures), and \UPDATED{(5) the test case does not distinguish between different error codes (i.e, it simply verifies that an error code is generated)}. Addressing such shortcomings make test cases more useful for root cause analysis.

\subsection{RQ3 - Feasibility}

The feasibility of data-driven mutation analysis depends on the required manual effort, which includes defining fault model specifications and injecting probes into the SUT source code. Also, the overhead introduced at runtime by the execution of the mutation operations may introduce delays in real-time systems and consequently cause failures. Finally, feasibility also depends on the overall duration of the mutation analysis process. 

To discuss manual effort, we measured, for each subject, (1) the number of \UPDATED{rows in the fault model specifications, as they match the number of operators manually identified and configured by an engineer}, \REVIEW{B.14}{and (2) the number of lines of code (LoC) added to the source code of our subjects. Since the number of added lines of code depends on the number of fault models per case study, we report the ratio of LoC per fault model. 
The lines of code added to the  source code of our subjects
include invocations to function \emph{mutate} (see Section~\ref{sec:generateAPI}) and additional utility code such as exit handlers used to clear the fault models loaded into memory.}

To address the overhead, we measured the execution time taken by every passing test case when executed with the original software and with any of the mutants generated by the approach. We exclude failing test cases because they may bias the results (e.g., failing assertions may terminate a test case earlier, while test timeout failures are detected when a test case execution takes too long). To account for performance variations due to the varying load of our HPC, we executed every test case three times. For every test case, we then computed the overhead of every mutant as the difference between the average execution time obtained with the mutants and that with the original software. Since different subjects are characterized by different types of messages being exchanged, we discuss the distribution of such overhead among our subjects.

Last, to discuss the overall duration of the mutation analysis process, we report the average time taken to execute \UPDATED{the test cases selected by the approach for every mutant, across three runs.}

\subsubsection*{Results}

\begin{table}[tb]
\caption{Manual effort and execution time.}
\label{table:costs} 
\scriptsize
\centering
\begin{tabular}{|
@{\hspace{1pt}}p{16mm}@{\hspace{2pt}}|
@{\hspace{1pt}}>{\raggedleft\arraybackslash}p{13mm}@{\hspace{1pt}}|
@{\hspace{1pt}}>{\raggedleft\arraybackslash}p{18mm}@{\hspace{1pt}}|
@{\hspace{1pt}}>{\raggedleft\arraybackslash}p{8mm}@{\hspace{1pt}}||
@{\hspace{1pt}}>{\raggedleft\arraybackslash}p{10mm}@{\hspace{1pt}}|
@{\hspace{1pt}}>{\raggedleft\arraybackslash}p{14mm}|}
\hline
\textbf{Subject}&\textbf{Configured} &\textbf{Configured} &\textbf{LoC} &\multicolumn{2}{c|}{\textbf{Execution time}}\\
&\textbf{Operators}&\textbf{Operators / FM}&\textbf{ / FM}&\textbf{Original} &\textbf{\APPR}\\
\hline
\ADCS	& 142 & 14.20 & 6.10 & \multirow{3}{*}{8.34 [h]} & \NEWRES{947.17} [h]\\
\GPS    & 23  & 23.00 & 2.72 &   & \NEWRES{42.02} [h]\\
\PDHU	& 29 &  9.66   & 4.33 &   & \NEWRES{60.36} [h] \\
\hline
\PARAM	& \NEWRES{44} & \NEWRES{13.33} & \NEWRES{7.64}& \NEWRES{0.015 [h]} & \NEWRES{0.12 [h]}  \\

\hline

\end{tabular}
\end{table}

\emph{Manual effort.} 
The left part of Table~\ref{table:costs} reports the measures related to manual effort. The number of operators configured per subject varies from 23 to 142, with an average between 9.66 and 23 operators per fault model. In our experiments, on average, it took between five and ten minutes to configure an operator \TSEmin{2.3}{(i.e., to read the paragraph of the software specifications related to a specific data item and to write the configuration values for an operator in the \APPR fault model).}
Given that the definition of test cases for safety-critical CPS components, such our case study subjects, takes days to complete, our industry partners found the required effort acceptable.
The same considerations hold for the number of LoC per fault model, whose average across subjects varied between 2.72 and 7.64\footnote{The exit handler includes one call for each fault model and thus subjects with less fault models show a lower average. For \PARAM, the larger number of lines of code is due to the need for recomputing a message checksum after the invocation of function \emph{mutate}.}. \TSEmin{1.3}{In total, the configuration of \APPR (i.e., configure operators and insert mutation probes) for \ADCS, \GPS, \PDHU, and \PARAM took around 20, 3, 4, and 6 working hours, respectively.}

\emph{Overhead.} Excluding outliers (i.e., values above $90^{th}$ percentile), the maximum execution overhead for 
\ADCS, \GPS, \PDHU, and \PARAM is 1.47\%, 3.16\%, 1.7\%, and \NEWRES{1.3\%}, respectively.
\REVIEW{D.7}{We did not observe any failure due to violated timing constraints (i.e., constraints defined within SUT test suites that verify both functional and time requirements); this indicates that \APPR does not introduce significant delay.}
Therefore, overall, the small overhead incurred by the subjects is acceptable and does not prevent the application of \APPR to real-time CPSs.

The right part of Table~\ref{table:costs} shows the \APPR analysis time. Although it is much larger than the execution times of test suites for the original SUTs, it is practically feasible. Indeed, in the worst case (i.e., \ADCS), mutation analysis can be performed in \NEWRES{9} hours with 100 parallel computation nodes; in safety-critical contexts, where development entails large costs, buying computation time on the Cloud is affordable. \UPDATED{Code-driven mutation analysis for systems with similar characteristics lasts considerably more~\cite{Ramler2017,Oscar:MASS:TSE}. For ESAIL and \PARAM, for example,  mutants sampling\footnote{Mutants sampling consists of selecting a subset of mutants to compute the mutation score~\cite{Guizzo:2020}. In our previous work~\cite{Oscar:MASS:TSE}, we have demonstrated that a fixed-width sequential confidence interval approach guarantees the accurate estimation of the mutation score.} makes code-driven mutation analysis feasible with parallel computation nodes, leading to 1800 hours (ESAIL) and 3 hours (\PARAM) of execution time, which is still significantly higher than the time required by \APPR. Without mutants sampling, mutation analysis of the whole ESAIL software may take up to 589,000 hours. We will evaluate the applicability of mutants sampling to \APPR in future work; however, although useful to derive an accurate mutation score, mutants sampling prevents the complete identification of test suite shortcomings. \TSEmin{1.5}{For this reason, we aim to study how mutants and test cases can be prioritized to generate accurate mutation analysis results while executing a subset of the mutants with a subset of the test cases.}}

\subsection{Data Availability}

The source code of our toolset, usage examples, and the data collected in our empirical evaluation are available~\cite{REPLICABILITY}. Our case study software cannot be provided since it is proprietary.

\subsection{Discussion}

Our results demonstrate that data-driven mutation analysis, as implemented in the \APPR approach, can effectively identify test suite shortcomings. Based on our results, \APPR does not lead to false positives, requires limited manual effort, and does not significantly affect the real-time properties of the software. Also, the overall execution time of the mutation analysis process is much lower than that of code-driven mutation analysis, which is the most similar approach proposed in the literature.

Our industry partners confirmed the correctness and usefulness of our results, as well as the practical feasibility of the approach.
Further, the effectiveness of the approach is particularly encouraging given that our case study subjects 
are safety-critical CPS software systems (i.e., satellite software components)
that were---before we started our experiments---already well tested based on several test suite inspection and validation activities, as per ECSS standards~\cite{ecss40C,ecss80C}.

To discuss the ease of adoption of \APPR in industry, we also need to consider the cost for the manual configuration of the approach (i.e., the definition of fault models and the insertion of probes into the SUT source code). \REVIEW{C.4}{Overall, setting up \APPR requires an amount of effort that is comparable to configuring other software testing and mutation analysis tools. For example, to generate unit test cases with KLEE~\cite{cadar2008klee}, which is a well-known test generation tool based on symbolic execution, it is necessary to implement test templates distinguishing the outputs and the inputs to be treated symbolically for each function under test, which may entail substantial effort for large software systems. Also, to perform code-driven mutation analysis with MASS~\cite{Oscar:MASS:TSE}, which is our recent code-driven mutation analysis tool for CPS software, it is necessary to update the configuration and execution scripts for the SUT (e.g., to collect code coverage data, compile with multiple compiler optimization options, determine test timeouts, and track test failures), an activity that requires a few hours to complete. Further, other mutation analysis tools require manual effort to be executed; for example, MART~\cite{MART} and Mull~\cite{denisov2018mull} require the software to be compiled with LLVM~\cite{LLVM}, which implies modifying compilation scripts and often leads to compilation errors that are hard to solve for large CPS software with many dependencies on third party libraries~\cite{Oscar:MASS:TSE}.} 
\REVIEW{C.3}{Finally, note that our implementation of \APPR enables engineers to introduce probes into the SUT source code using dedicated annotation labels (i.e., specific comment keywords); such annotations do not alter the execution of production code but can be used to automatically perform mutation analysis within a CI/CD pipeline without requiring manual intervention.}

 Compared to traditional code-driven mutation analysis, data-driven mutation analysis may still require more manual effort and end-user expertise (e.g., knowledge of the SUT) because the former is usually performed automatically by software toolsets without the need for a manually specified fault model. However,  data-driven mutation analysis provides more readily interpretable results than code-driven mutation analysis, thanks to our three mutation analysis metrics. For example, the lack of coverage for a FVAT mutation operation applied to a specific data item indicates that the test suite does not exercise a value above the FVAT threshold parameter for that data item (e.g., it does not exercise non-nominal cases); in the case of code-driven mutation analysis, it is often difficult to determine which missing test input prevents the test suite from killing a mutant (e.g, first, engineers need to determine if the mutant is nonequivalent to the original program, which is in itself complicated). Finally, based on our experience, data-driven mutation analysis leads to a lower number of mutants than traditional code-driven mutation analysis. For example, in the case of \PARAM, \APPR generates 44 mutants while code-driven mutation analysis, based on our previous work~\cite{Oscar:MASS:TSE}, leads to 3931 mutants (346, if mutants sampling is applied). The low number of mutants generated by \APPR leads to a low number of mutants to be inspected (e.g., because not detected by the test suite), which greatly reduces the cost of the analysis. For example, in the case of \PARAM, with \APPR, we inspected only 4 mutants, which sharply contrasts with code-driven mutation analysis that led, in our previous studies, to examine 1179 mutants (or 50 with mutants sampling~\cite{Oscar:MASS:TSE}). 

Based on the above, we can conclude that the manual effort required by \APPR is comparable or lower than that required by code-driven mutation analysis, which is already adopted in certain contexts~\cite{MTFacebook,MTGoogle}, and other well-known software testing tools. Consequently, we believe that data-driven mutation analysis can be realistically adopted in the CPS industry, given that it targets problems that are particularly relevant for CPS software. 

\TSEmin{1.6}{Finally, we report the results of an independent evaluation of code-driven and data-driven mutation analysis performed by our industry partners in the context of the ESA project supporting this research. According to our industry partners, the percentage of critical test suite shortcomings detected by code-driven mutations is 38.24\%; for data-driven mutation, critical shortcomings add up to 57\%. Examples of critical shortcomings include message types not being exercised, input partitions not being covered, or test suites not detecting the presence of unexpected non-nominal data values. These results further highlight the potential of data-driven mutation analysis.}

\subsection{Threats to validity}

\emph{External validity}. We have selected industrial CPSs of diverse size, tested with different types of test suites. They are developed according to space safety standards and are thus  representative of CPS software adhering to safety regulations. Also, \ESAIL is larger than any other industrial system considered in the mutation analysis literature to date~\cite{Ramler2017,delgado2018evaluation,Baker2013,denisov2018mull}.

\UPDATED{\emph{Internal validity}. To minimize implementation errors, we have extensively tested our toolset; we provide both the test cases and the \APPR source code. Also, the \APPR toolset has been developed in the context of an ESA project which required following ECSS practices for category D software validation and documentation.}

\UPDATED{\emph{Construct validity}. The indicators selected for cost estimation (configured operators and LoC) are directly linked to the activities of the end-user and are thus appropriate.
\REVIEW{B.14}{We rely on LOC per Fault Model since LOC is an objective though imperfect indicator of effort. Ideally, though far from being straightforward, effort discussions should also be based on data collected from practitioners. However, since the reported LOC/FM values are very small, there is not much room for interpretation in our context.}  
We therefore leave empirical studies with human subjects to future work. 
\TSEmin{1.3, 2.3}{Future studies should evaluate not only the effort required to configure \APPR operators but also the effort required to determine what data to mutate; however, based on our experience, determining what to mutate is an easy task in safety-critical CPS because the software specification documents provide a clear description of what are the communicating components, the type of messages being exchanged, and the message structure. Further, future studies should also report on the likelihood of equivalent mutants caused by human errors; for example, operators can be misconfigured (e.g., the misconfigured operator may generate values outside the representable range that will simply not be transmitted over the communication channel) or probes can be inserted in the wrong location (e.g., inserting the mutation probe in a block of code that is executed after the data to be mutated has been processed by the SUT). Such mistakes may lead to false alarms that are noticed only when investigating uncovered fault models, uncovered mutation operators, and live mutants; after fixing such mistakes, the engineers need to re-execute the test suite with the mutants generated by the subset of operators that have been reconfigured.}
Though user studies are required to confirm this statement, our industry partners found the effort entailed by the approach to be justified by its benefits (e.g., they identified message types not being tested). 
}

\section{Conclusion}
\label{sec:conclusion}

Assessing the quality of test suites is necessary 
in the case of large, safety-critical systems, such as most CPSs, where software failures may lead to severe consequences including loss of human lives or environmental damages. This problem is made even more acute due to the fact that such test suites are usually manually constructed. 

In this paper, we have introduced data-driven mutation analysis, a new paradigm and associated techniques to assess the effectiveness of a test suite in detecting interoperability faults. Interoperability faults are a major source of failures in CPSs but they are not targeted by existing mutation analysis approaches.
Data-driven mutation analysis works by modifying (i.e., mutating) the data exchanged by software components during test execution. Mutations are performed through a set of mutation operators that are configured according to a fault model provided by software engineers, following a proposed methodology. 
Test suite assessment is based on three metrics: (1) fault model coverage, (2) mutation operation coverage, (3) and covered mutation score. These metrics enable engineers to determine specific weaknesses in test suites (e.g., oracles with low covered mutation score).

We have proposed a partially automated technique, \APPR, that is applicable to a vast range of systems since it mutates the data exchanged through data buffers, by relying on a tabular fault model that can be inexpensively loaded into memory.
Based on interactions with engineers, we have defined a set of mutation operators that enable the simulation of common data faults causing interoperability problems in CPSs.
We have provided a methodology that supports the definition of a fault model based on the nature, representation type, and dependencies of the data to mutate.
\REVIEW{C.9}{In future work, additional mutation operators can be defined to address specific aspects of other data structures, e.g., changes to structures of trees or lists.} %
\REVIEW{D.7}{Furthermore, we will assess mutant sampling strategies to reduce execution time without dampening effectiveness.}

We empirically evaluated \APPR by applying it to test suites of commercial CPS components in the space domain that are currently deployed on orbit. Our results show that \APPR can identify a large and diverse set of test suite shortcomings, entails limited modelling and execution costs, and is not affected by redundant and equivalent mutants.

\ifCLASSOPTIONcompsoc
  \section*{Acknowledgments}
\else
  \section*{Acknowledgment}
\fi

This work has been funded by the European Space Agency (ITT-1-9873/FAQAS),
the European Research Council (ERC) under the European Union’s Horizon 2020 research and innovation programme (grant agreement No 694277), 
and NSERC Discovery and Canada Research Chair programs. Authors would like to thank the ESA ESTEC officers, the \EduardoSpace team, 
\YagoSpace software engineers, and \YAGO{} for their valuable support.

\ifCLASSOPTIONcaptionsoff
  \newpage
\fi

\bibliographystyle{./bibliography/IEEEtran}
\bibliography{./bibliography/dataMutationTesting}

\begin{IEEEbiography}[{\includegraphics[width=1in,height=1.25in,clip,keepaspectratio]{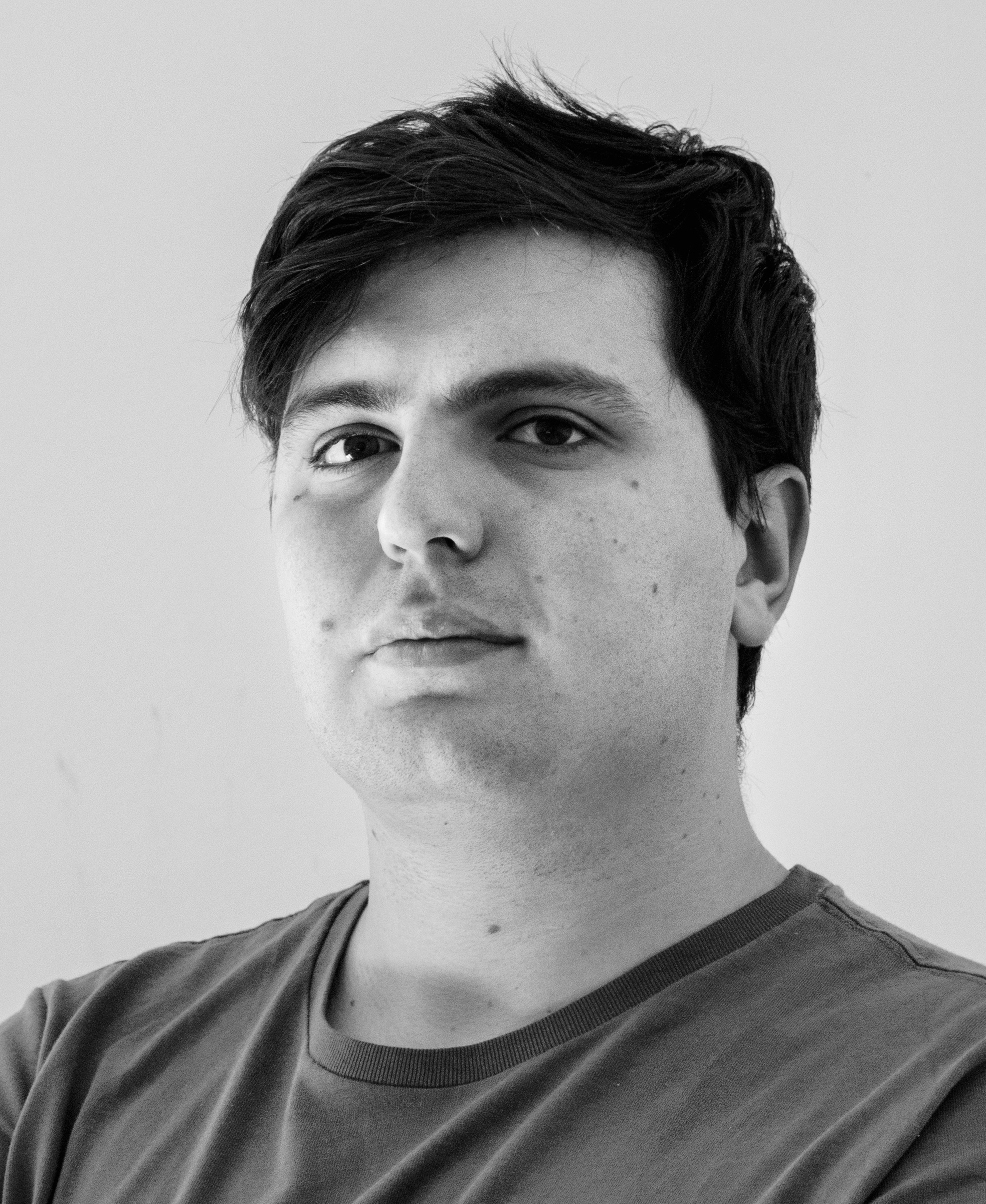}}]{Enrico Viganò} 
is a Research and Development Specialist at the Interdisciplinary Centre for Security, Reliability and Trust (SnT), University of Luxembourg. 

His research interests lie in automated testing for space software. He is currently involved in projects with the European Space Agency and industry partners from the space domain. 
\end{IEEEbiography}

\begin{IEEEbiography}[{\includegraphics[width=1in,height=1.25in,clip,keepaspectratio]{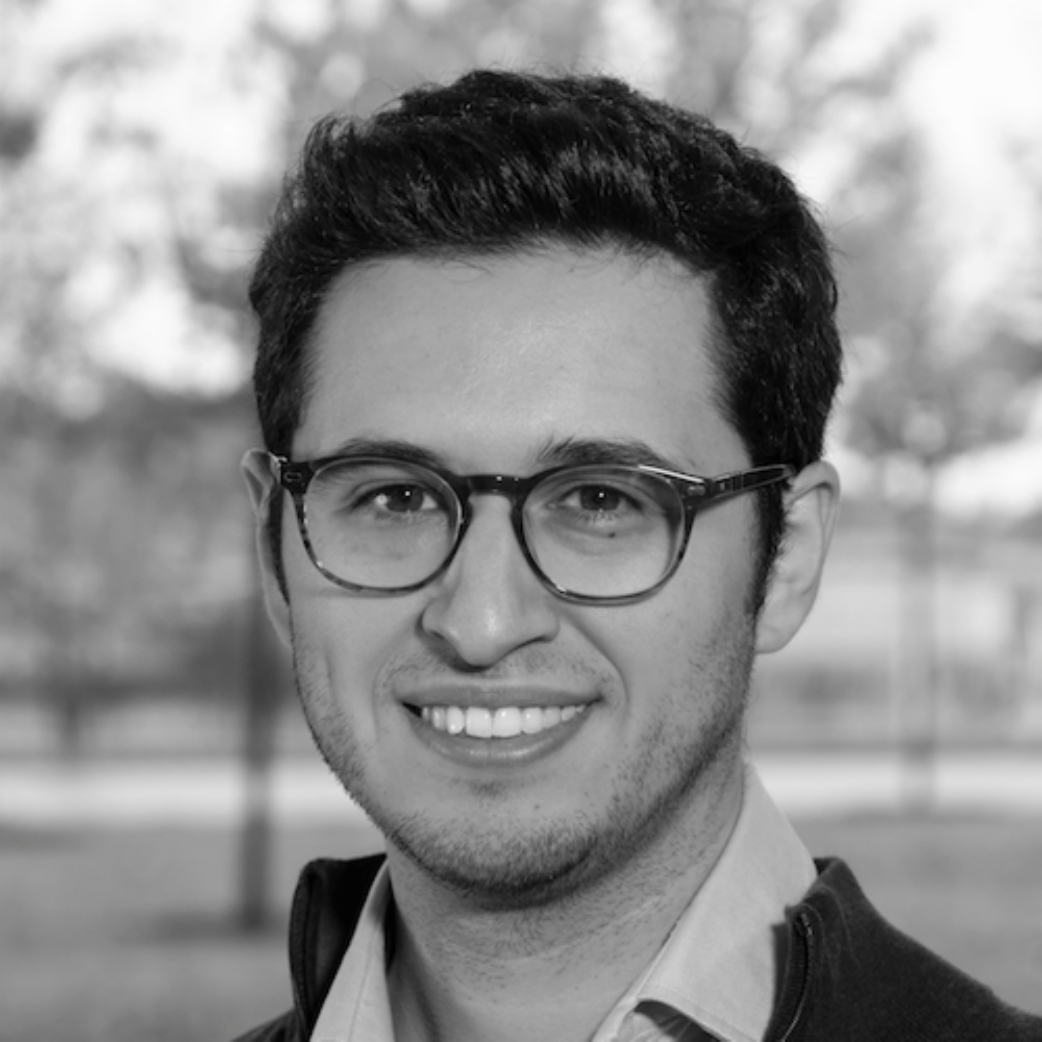}}]{Oscar Cornejo} 
is Senior Software QA Engineer at SES Luxembourg. Previously, Oscar had been a Research Associate at the Interdisciplinary Centre for Security, Reliability and Trust (SnT), University of Luxembourg. He obtained his PhD degree in Computer Science from the University of Milano - Bicocca in 2019.

His interests are in software engineering, focusing on automated software testing and program analysis. He is currently working on R\&D projects in the space domain. 
\end{IEEEbiography}

\begin{IEEEbiography}[{\includegraphics[width=1in,height=1.25in,clip,keepaspectratio]{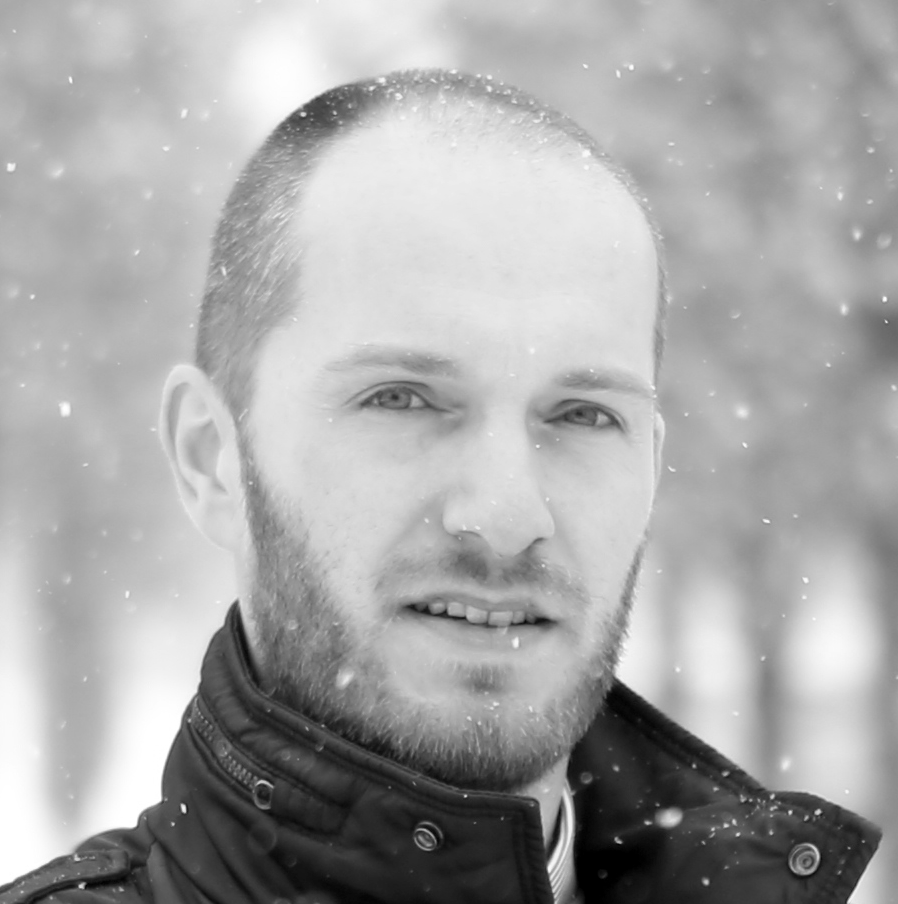}}]{Fabrizio Pastore}
is Chief Scientist II at the Interdisciplinary Centre for Security, Reliability and Trust (SnT), University of Luxembourg. He obtained his PhD in Computer Science in 2010 from the University of Milano - Bicocca.

His research interests concern automated software testing, including security testing and testing of AI-based systems; his work relies on the integrated analysis of different types of artefacts (e.g., requirements,  models, source code, and execution traces). He is active in several industry partnerships and national, ESA, and EU-funded research projects.
\end{IEEEbiography}

\begin{IEEEbiography}[{\includegraphics[width=1in,height=1.25in,clip,keepaspectratio]{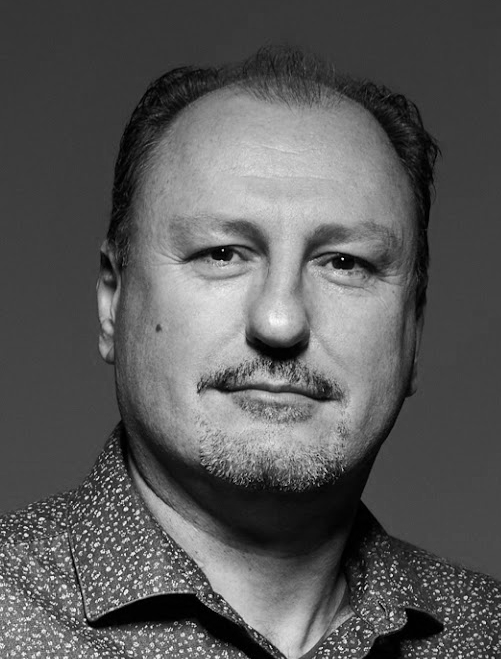}}]{Lionel C. Briand} is professor of software engineering and has shared appointments between (1) School of Electrical Engineering and Computer Science, University of Ottawa, Canada and (2) The SnT centre for Security, Reliability, and Trust, University of Luxembourg. He is the head of the SVV department at the SnT Centre and a Canada Research Chair in Intelligent Software Dependability and Compliance (Tier 1).

He has conducted applied research in collaboration with industry for more than 25 years, including projects in the automotive, aerospace, manufacturing, financial, and energy domains. In 2016, he received an ERC Advanced grant, the most prestigious European individual research award. He was elevated to the grades of IEEE and ACM fellow, granted the ACM SIGSOFT Outstanding Research Award (2022), the IEEE Computer Society Harlan Mills award (2012), and the IEEE Reliability Society Engineer-of-the-year award (2013) for his work on software verification and testing. 
His research interests include: Testing and verification, search-based software engineering, model-driven development, requirements engineering, and empirical software engineering.
\end{IEEEbiography}

\end{document}